# Anisotropic conductivity of uncharged domain walls in BiFeO$_3$

Anna N. Morozovska[1], Rama K. Vasudevan[2], Peter Maksymovych[3],

Sergei V. Kalinin[3,*] and Eugene A. Eliseev[4,†]

[1] Institute of Physics, National Academy of Sciences of Ukraine,

46, pr. Nauki, 03028 Kiev, Ukraine

[2] School of Materials Science and Engineering, University of New South Wales, Kensington

Sydney 2052 Australia

[3] Center for Nanophase Materials Science,

Oak Ridge National Laboratory, Oak Ridge, TN, 37831

[4] Institute for Problems of Materials Science, National Academy of Sciences of Ukraine,

3, Krjijanovskogo, 03142 Kiev, Ukraine

## Abstract

Experimental observations suggest that nominally uncharged, as-grown domain walls in ferroelectric thin films can be conductive, yet comprehensive theoretical models to explain this behavior are lacking. Here, Landau theory is used to evolve an analytical treatment of the anisotropic carrier accumulation by nominally uncharged domain walls in multiferroic BiFeO$_3$. Strong angular dependence of the carrier accumulation by 180-degree domain walls originates from local band bending via angle-dependent electrostriction and flexoelectric coupling mechanisms. Theoretical results are in qualitative agreement with experimental data, and provide a Landau-Ginzburg-Devonshire counterpart that is consistent with recent first principles calculations. These studies suggest that a significantly more diverse range of domain wall structures could possess novel electronic properties than previously believed. Similarly, emergent electronic behaviors at ferroic walls are typically underpinned by multiple mechanisms, necessitating first-principle studies of corresponding coupling parameters.

[*] Corresponding author e-mail sergei2@ornl.gov

[†] Corresponding author e-mail eugene.a.eliseev@gmail.com

**Introduction**

Functionality of condensed matter systems are often controlled by small local distortions from ideal structure traditionally described through order parameters, which can represent any of several energetically degenerate ground states. In this description, regions with the same order parameter values (i.e. domains) are separated by boundaries, classified as topological defects [1], which possess different symmetry and can display markedly different and new properties not exhibited in the bulk material [2, 3]. Examples include ferroelectric, ferroelastic and magnetic domain walls in ferroic materials, as well as more subtle distortions such as Jahn-Teller walls [4]. From a technological perspective, ferroelectric and ferroelastic domain walls have been a foci of interest due to strong coupling with lattice strain, and correspondingly, significant strain-mediated effects such as polarization rotations [5], and ferroelastic phase transitions at the interface [6]. Additionally, the discovery of novel electronic properties arising at these domain walls has catalysed significant experimental and theoretical interest in the topic [4-6, 7, 8].

It has been known for decades that the existence of charged domain walls in ferroelectric semiconductors should result in accumulation or inversion on adjacent sides of the domain wall, and should thus lead to enhanced conductivity at these sites [9]. Remarkably, experimental verification of this prediction occurred only recently, by Seidel *et al* who used scanning probe microscopy (SPM) methods to report room-temperature metallic conductivity of 180° and 109° domain walls in BiFeO$_3$ [10, 11]. In their report, the authors found that the 71° domain walls did not exhibit conductivity. However, recent investigations of nominally uncharged fabricated vortex structures in BFO show an order of magnitude increase in conductivity over single domain regions [12]. Other experiments have shown that nominally uncharged as-grown 71° domain walls can also exhibit enhanced conductivity [13]. Farokhipoor *et al* [13] reported that the conductivity at the domain walls in thin (40-70nm) BFO films was similar for the 71° and 109° as-grown domain walls, suggesting strain-related effects, as opposed to bound charge at the domain wall, are an important (yet poorly understood) factor in determining local electronic properties.

**1. Domain wall conductance in BiFeO$_3$**

Ferroelectric bismuth ferrite, BiFeO$_3$ (BFO), is the one of the most promising multiferroics due to giant spontaneous polarization (about 0.9 C/m$^2$ at room temperature), high ferroelectric Curie temperature (~1100 K), strong antiferromagnetism (magnetic Curie temperature ~650 K) and pronounced structural ordering (oxygen octahedron tilt) [14]. Distorted perovskite-structured BiFeO$_3$ also exhibits intriguing physical properties, which most probably originate from the complex interplay between coexisting structural, polar and magnetic ordering in the single domain regions and especially at the domain walls [14, 15], where the electronic properties can changes drastically [10, 11, 12, 13].

To illustrate variability of the conductivity response at BiFeO$_3$ domain walls, shown in **Figure 1** are experimental studies with current-AFM (c-AFM) and Piezoresponse Force Microscopy (PFM) of a 200nm BFO film grown on DyScO$_3$ substrate. The film consists of a 71° in-plane striped domain structure, with a domain spacing of ~200 nm [16]. To form the other two types of domain walls, a square area was poled with the tip held at -6V. This results in both ferroelectric (180°) and ferroelastic (71°) switching, as shown by the Vertical and Lateral PFM phase images shown as insets in **Fig. 1a**. The current-AFM (c-AFM) image, taken with the tip held at voltage V = -2.6V, is also shown in **Fig. 1a**, and indicates that conduction can be observed at all three types of domain walls. A close up of the top-left quadrant is shown in **Fig. 1b**, where a conducting 180° wall segment can be seen. Additionally, ultra-high vacuum (UHV) SPM studies of the film, shown in the topography and c-AFM image in **Fig. 1c,d** indicate that the as-grown 71° domain walls, which are nominally uncharged, are conducting. Collectively these experimental findings suggest that even uncharged domain walls and topological defects can exhibit electric conduction, seemingly at odds with existing theories as to the origins of the static domain wall's conductivity.

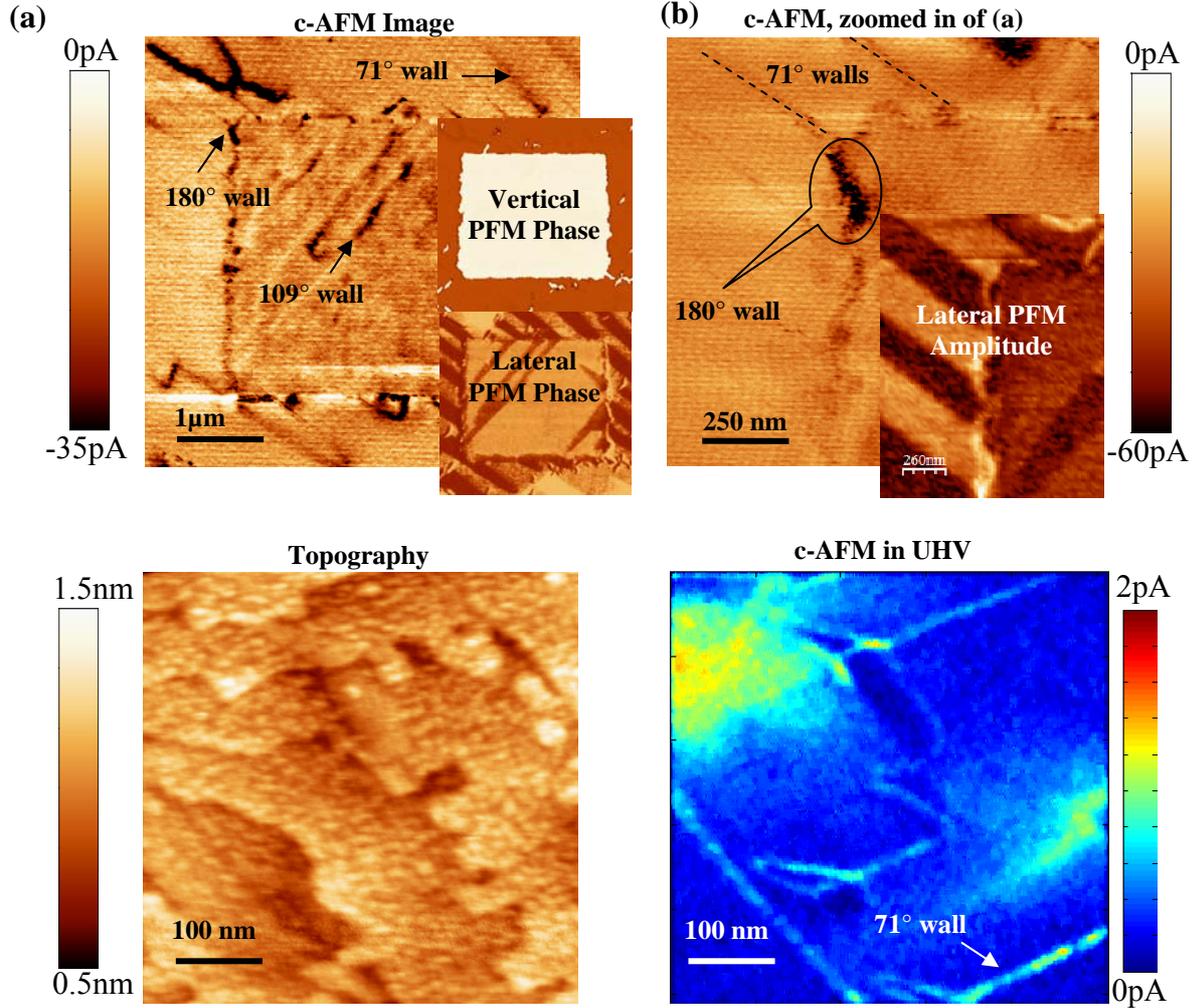

**Figure 1. (a)** c-AFM taken with $V_{tip}$ = -2.6V reveals that the 109° and 71° uncharged domain walls in BFO are conducting. The dark corner segment is a 180° wall, as indicated. Vertical and lateral PFM phase images are shown inset. **(b)** Zoomed part of (a), showing the conductive 180° wall segment, as well as 71° walls. Lateral PFM amplitude is shown on the inset. **(c)** Topography and **(d)** Simultaneous c-AFM image, taken with $V_{sample}$ = +2V, in UHV of as-grown, conducting 71° domain walls.

It is generally accepted that uncharged walls are thermodynamically stable and most energetically preferable in bulk ferroelectrics [17]. Consequently the walls can be readily created in required amount and their spatial location can be manipulated and controlled with nanoscale resolution by e.g. SPM [10, 11, 12]. While the charge state of the domain walls in realistic systems is generally unknown, it is of interest to explore, theoretically, the electronic conductivity properties of uncharged walls at a prototypical ferroic wall. .

Ferroelectric, structural and magnetic properties of BiFeO$_3$ are relatively well-studied both experimentally (see e.g. [18]) and theoretically (see e.g. [19]). Numerical *ab initio* calculations [20] and phase-field modelling based [21] on phenomenological Landau theory are available. Electronic properties of the domain walls, on the other hand, are much less studied. First-principles calculations of ferroelectric domain walls in BiFeO$_3$ [20] showed that band gap narrows on the value 0.2 eV, 0.1 eV and 0.05 eV at nominally uncharged 180-, 109- and 71-degree domain walls correspondingly. Despite this, analytical Landau-type theory of the uncharged domain walls' conductivity in rhombohedral ferroelectrics is currently absent. For justice it is worth noting that Landau theory is a powerful method that has proven capable of predicting charged domain walls' static conductivity in ferroelectric-semiconductors in 1969 [22], the conductivity mechanism stemming from compensation of polarization charge discontinuity by mobile carriers in the material. Analytical Landau-type theory was further developed for charged walls in uniaxial [23, 24] and multiaxial tetragonal ferroelectrics [25], improper ferroelectrics [26] and twin walls in incipient ferroelectrics – ferroelastics [27]. These studies agree with recent experiments on Pb(Zr,Ti)O$_3$ [28, 29], ErMnO$_3$ [30] and LiNbO$_3$ [31].

The gap between experimental observations of conductivity at nominally uncharged walls and the existing theoretical frameworks, as well as the potential for use of such walls in technological applications ('domain wall nanoelectronics' [32]), motivate an analytical study of the free carrier accumulation by nominally uncharged 180-, 109- and 71-degree domain walls in bulk BiFeO$_3$. Using Landau-Ginzburg-Devonshire theory it is shown that polarization changes via flexoelectric coupling and inhomogeneous elastic strains give rise to strongly anisotropic local band bending and carrier accumulation by uncharged domain walls. These results rationalize experimentally observed conductivity at nominally uncharged domain walls, and further suggest that spatial modulation of conduction via secondary strain-related effects is possible even in systems where charged domain walls are thermodynamically unstable.

**2. Basic equations and their analytical solution**

Here, the conductivity of uncharged domain walls in BiFeO$_3$ is explored. For rhombohedral materials, the three types of domain walls are 180-, 109- and 71-degree walls, as shown in **Figs.2**(a-c), respectively.

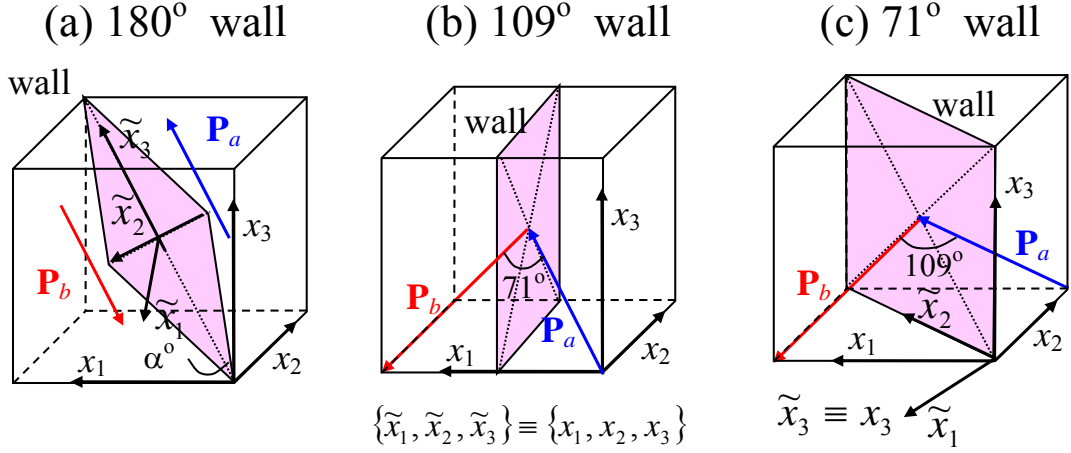

**Figure 2.** Rotated coordinate frame $\{\tilde{x}_1, \tilde{x}_2, \tilde{x}_3\}$ choice for (a) 180-degree, (b) 109-degree and (c) 71-degree uncharged domain walls in a rhombohedral ferroelectric BiFeO$_3$. Pseudo-cubic crystallographic axes are $\{x_1, x_2, x_3\}$. Rotation angle α is counted from $\tilde{x}_3$ axes.

Within Landau-Ginzburg-Devonshire theory, polarization-dependent elastic strains variations $\delta u_{ij}$ caused by uncharged domain walls have the form (see **Suppl. Materials**):

$$\delta \tilde{u}_{22} = \delta \tilde{u}_{33} = \delta \tilde{u}_{23} = 0, \tag{1a}$$

$$\begin{aligned}\delta \tilde{u}_{i1} &= \vartheta_{i1}\left(\tilde{P}_1^2 - \left(\tilde{P}_1^S\right)^2\right) + \vartheta_{i2}\left(\tilde{P}_2^2 - \left(\tilde{P}_2^S\right)^2\right) + \vartheta_{i3}\left(\tilde{P}_3^2 - \left(\tilde{P}_3^S\right)^2\right) + \vartheta_{i4}\left(\tilde{P}_2\tilde{P}_3 - \tilde{P}_2^S\tilde{P}_3^S\right) \\ &+ \vartheta_{i5}\left(\tilde{P}_1\tilde{P}_3 - \tilde{P}_1^S\tilde{P}_3^S\right) + \vartheta_{i6}\left(\tilde{P}_1\tilde{P}_2 - \tilde{P}_1^S\tilde{P}_2^S\right) + \Psi_{i1}\frac{\partial\tilde{P}_1}{\partial\tilde{x}_1} + \Psi_{i2}\frac{\partial\tilde{P}_2}{\partial\tilde{x}_1} + \Psi_{i3}\frac{\partial\tilde{P}_3}{\partial\tilde{x}_1}\end{aligned} \tag{1b}$$

.Here $\delta u_{ij}(\mathbf{r}) = u_{ij}(\mathbf{r}) - u_{ij}^S$, where $u_{ij}^S$ is the spontaneous strain; $i = 1,2,3$, tilda "~" defines polarization vector $\tilde{P}_i$, its spontaneous value $\tilde{P}_i^S$ and tensor components in the coordinate frame $\{\tilde{x}_1, \tilde{x}_2, \tilde{x}_3\}$ rotated with respect to the pseudo-cubic crystallographic axes $\{x_1, x_2, x_3\}$ as shown in **Figs.2a-c.** Coordinates transformation for 180-degree uncharged domain walls in rhombohedral phase is:

$$\tilde{x}_1 = \sqrt{\frac{2}{3}}\sin\alpha\, x_1 + x_2\left(\frac{\cos\alpha}{\sqrt{2}} - \frac{\sin\alpha}{\sqrt{6}}\right) - x_3\left(\frac{\cos\alpha}{\sqrt{2}} + \frac{\sin\alpha}{\sqrt{6}}\right), \tag{2a}$$

$$\tilde{x}_2 = -\sqrt{\frac{2}{3}}\cos\alpha\, x_1 + x_2\left(\frac{\cos\alpha}{\sqrt{6}} + \frac{\sin\alpha}{\sqrt{2}}\right) + x_3\left(\frac{\cos\alpha}{\sqrt{6}} - \frac{\sin\alpha}{\sqrt{2}}\right) \text{ and } \tilde{x}_3 = \frac{x_1 + x_2 + x_3}{\sqrt{3}}, \tag{2b}$$

where α is the rotation angle. Coordinates transformation for 109-degree and 71-degree domain walls in rhombohedral phase is

$$\tilde{x}_1 = x_1 \cos\alpha - x_2 \sin\alpha, \quad \tilde{x}_2 = x_1 \sin\alpha + x_2 \cos\alpha, \quad \tilde{x}_3 = x_3, \qquad (2c)$$

To maintain charge neutrality, the rotation angle α can be arbitrary for 180-degree uncharged domain walls in the equilibrium, while $\alpha = 0, \pi$ for 109-degree and $\alpha = -\pi/4, 3\pi/4$ for 71-degree uncharged domain walls (see **Table 1**).

Coefficients $\vartheta_{ij}$ in Eqs.(1) are proportional to the combinations of the electrostriction tensors' coefficients $Q_{ij}$ and $\tilde{Q}_{ij}$, and elastic compliances $s_{ij}$ and $\tilde{s}_{ij}$ defined in the crystallographic $\{x_1, x_2, x_3\}$ and rotated $\{\tilde{x}_1, \tilde{x}_2, \tilde{x}_3\}$ frames correspondingly. Coefficients $\Psi_{ij}$ are proportional to the flexoelectric coupling coefficients $F_{ij}$ and $\tilde{F}_{ij}$, $s_{ij}$ and $\tilde{s}_{ij}$. The coefficients and rotated tensors' components are listed in the **Supplemental Materials, Tables S1.** The **Table S1a** is valid for uncharged 180-degree domain walls rotated at angle α, Voigt notations are used. The **Table S1b** is valid for uncharged 109- and 71-degree domain walls in rhombohedral ferroelectric phases.

In Equations (1) the values of spontaneous polarization components $\tilde{P}_i^S$ depend on the wall type. Polarization components $\tilde{P}_i$, which depend on the distance $\tilde{x}_1$ from the domain wall plane $\{\tilde{x}_2, \tilde{x}_3\}$, can be calculated numerically from the system of coupled LGD-type equations allowing for electrostriction, flexoelectric coupling and depolarization field $\tilde{E}_1^d(\tilde{x}_1)$ acting on the component $\tilde{P}_1(\tilde{x}_1)$ (see details in Ref. [25]). Due to the smallness of flexoelectric coupling, it is possible to use a perturbation approach in order to derive an accurate analytical expression for $\tilde{P}_i(\tilde{x}_1)$ for considered domain wall geometries. Approximate analytical expressions for polarization components $\tilde{P}_i(\tilde{x}_1)$ are listed in the **Table 1.**

**Table 1.** Approximate analytical expressions for polarization components $\tilde{P}_i(\tilde{x}_1)$ in the vicinity of uncharged domain walls in BiFeO$_3$

| Domain wall (DW) | 109-degree | 71-degree | 180-degree |
|---|---|---|---|
| **Rotation angle α** | $\alpha = 0, \pi$ in the **rhombohedral** phase DW is absent in the **tetragonal** phase | $\alpha = -\pi/4, 3\pi/4$ in the **rhombohedral** phase DW is absent in the **tetragonal** phase | α is arbitrary in both **rhombohedral** and **tetragonal** phases |
| **Component $\tilde{P}_3$** | $\tilde{P}_3^S \tanh\left(\dfrac{\tilde{x}_1}{L_c}\right)$ | $\tilde{P}_3^S \tanh\left(\dfrac{\tilde{x}_1}{L_c}\right)$ | $\tilde{P}_3^S \tanh\left(\dfrac{\tilde{x}_1}{L_c}\right)$ |
| **Component $\widetilde{P}_2$** ** | $\widetilde{P}_2^S \tanh\left(\dfrac{\widetilde{x}_1}{L_c}\right)$ | 0 | $P_b \cosh^{-2}\left(\dfrac{\widetilde{x}_1}{L_c}\right) + \dfrac{f_2^Q}{2\beta}\dfrac{\partial \widetilde{P}_3^2}{\partial \widetilde{x}_1}$ $+ \dfrac{q_2}{2\beta}\widetilde{P}_3\left(\left(\widetilde{P}_3^S\right)^2 - \widetilde{P}_3^2\right)$ |
| **Component $\widetilde{P}_1$** | $\widetilde{P}_1^S +$ $\dfrac{\varepsilon_0\varepsilon_b f_1^Q}{1+2\beta\varepsilon_0\varepsilon_b}\left(\dfrac{\partial \widetilde{P}_3^2}{\partial \widetilde{x}_1} + \dfrac{\partial \widetilde{P}_2^2}{\partial \widetilde{x}_1}\right)$ | $\widetilde{P}_1^S + \dfrac{\varepsilon_0\varepsilon_b f_1^Q}{1+2\beta\varepsilon_0\varepsilon_b}\dfrac{\partial \widetilde{P}_3^2}{\partial \widetilde{x}_1}$ | $\dfrac{\varepsilon_0\varepsilon_b f_1^Q}{1+2\beta\varepsilon_0\varepsilon_b}\dfrac{\partial \widetilde{P}_3^2}{\partial \widetilde{x}_1} +$ $\dfrac{\varepsilon_0\varepsilon_b q_1 \widetilde{P}_3}{1+2\beta\varepsilon_0\varepsilon_b}\left(\left(\widetilde{P}_3^S\right)^2 - \widetilde{P}_3^2\right)$ |
| **Correlation length** | $L_c = \sqrt{g_{44}/(-2a_1)}$ | $L_c = \sqrt{g_{44}/(-2a_1)}$ | $L_c = \sqrt{\widetilde{g}_{66}/(-2a_1)}$ |
| **Constant $\beta$*** | $\beta \approx a_1 + \widetilde{a}_{12}\left(\widetilde{P}_2^S\right)^2 + a_{12}\left(\widetilde{P}_3^S\right)^2$ | $\beta \approx a_1 + a_{12}\left(\widetilde{P}_3^S\right)^2$ | $\beta \approx a_1 + \left(\widetilde{a}_{13} + \widetilde{Q}_{44}^2/2\widetilde{s}_{44}\right)\widetilde{P}_S^2$ |
| **LGD-expansion coefficients*** | $\widetilde{a}_{12} = a_{12} + 3\dfrac{2a_{11} - a_{12}}{2}\sin^2(2\alpha),$ $\widetilde{g}_{66} = g_{44} + \sin^2(2\alpha)\left(\dfrac{g_{11} - g_{12}}{2} - g_{44}\right)$ | | $\widetilde{a}_{12} = (2a_{11} + a_{12})/2,\ \widetilde{a}_{13} = 2a_{11},$ $\widetilde{g}_{66} = (g_{11} - g_{12} + 4g_{44})/6$ |
| **Components $\widetilde{P}_i^S$** | $\widetilde{P}_1^S \neq 0,$ $\widetilde{P}_2^S = \pm\widetilde{P}_3^S \neq 0$ | $\widetilde{P}_1^S \neq 0,\ \widetilde{P}_2^S \equiv 0,$ $\widetilde{P}_3^S \neq 0$ | $\widetilde{P}_1^S \equiv 0,\ \widetilde{P}_2^S \equiv 0,\ \widetilde{P}_3^S \neq 0$ |
| **Constants $f_1^Q$ and $f_2^Q$** | 109- and 71-degree walls in rhombohedral phase: $f_1^Q = \widetilde{F}_{12}\dfrac{Q_{11}s_{12} - s_{11}Q_{12}}{s_{11}\widetilde{s}_{11} - s_{12}^2} + F_{12}\dfrac{s_{12}Q_{12} - \widetilde{s}_{11}Q_{11}}{s_{11}\widetilde{s}_{11} - s_{12}^2},$ $f_2^Q = \widetilde{F}_{26}\dfrac{Q_{11}s_{12} - s_{11}Q_{12}}{s_{11}\widetilde{s}_{11} - s_{12}^2} \equiv 0$ (since $\widetilde{F}_{26}(0) = \widetilde{F}_{26}(\pm\pi/4) = 0$) 180-degree walls in rhombohedral phase: $f_1^Q = -\dfrac{\widetilde{F}_{13}\widetilde{Q}_{33}}{\widetilde{s}_{33}} - \dfrac{\widetilde{F}_{15}\widetilde{Q}_{15}\widetilde{s}_{44}\widetilde{s}_{33} + 2\left(\widetilde{F}_{12}\widetilde{s}_{44} + 2\widetilde{F}_{14}\widetilde{s}_{14} - \widetilde{s}_{13}\widetilde{s}_{44}\left(\widetilde{F}_{13}/\widetilde{s}_{33}\right)\right)\left(\widetilde{Q}_{33}\widetilde{s}_{13} - \widetilde{Q}_{13}\widetilde{s}_{33}\right)}{2\left(\widetilde{s}_{14}^2\widetilde{s}_{33} + \left(\widetilde{s}_{13}^2 - \widetilde{s}_{11}\widetilde{s}_{33}\right)\widetilde{s}_{44}\right)},$ $f_2^Q = \widetilde{F}_{15}\dfrac{4\widetilde{s}_{14}\left(\widetilde{Q}_{33}\widetilde{s}_{13} - \widetilde{Q}_{13}\widetilde{s}_{33}\right) + \widetilde{s}_{33}\left(\widetilde{Q}_{44}\widetilde{s}_{14} - \widetilde{Q}_{14}\widetilde{s}_{44}\right)}{2\left(\widetilde{s}_{14}^2\widetilde{s}_{33} + \left(\widetilde{s}_{13}^2 - \widetilde{s}_{11}\widetilde{s}_{33}\right)\widetilde{s}_{44}\right)}$ | | |
| **Constants $q_1$ and $q_2$** | 180-degree walls in rhombohedral phase: $q_1 = -\dfrac{\left(\widetilde{Q}_{33}\widetilde{s}_{13} - \widetilde{Q}_{13}\widetilde{s}_{33}\right)\widetilde{Q}_{15}\widetilde{s}_{44}}{\widetilde{s}_{14}^2\widetilde{s}_{33} + \left(\widetilde{s}_{13}^2 - \widetilde{s}_{11}\widetilde{s}_{33}\right)\widetilde{s}_{44}},\quad q_2 = \dfrac{\left(\widetilde{Q}_{33}\widetilde{s}_{13} - \widetilde{Q}_{13}\widetilde{s}_{33}\right)\left(\widetilde{Q}_{44}\widetilde{s}_{14} - \widetilde{Q}_{14}\widetilde{s}_{44}\right)}{\widetilde{s}_{14}^2\widetilde{s}_{33} + \left(\widetilde{s}_{13}^2 - \widetilde{s}_{11}\widetilde{s}_{33}\right)\widetilde{s}_{44}}.$ 109- and 71-degree walls in rhombohedral phase: $q_1 = q_2 = 0$ | | |

* Coefficients $a_1$, $a_{12}$ are LGD-potential expansion coefficients values in the crystallographic frame $\{x_1, x_2, x_3\}$, where $g_{ij}$ are the gradient coefficients [33]; $\varepsilon_0\varepsilon_b$ is the background dielectric constant.

Polarization variations $\delta\widetilde{P}_i(\widetilde{x}_1) \sim f_i^Q\left(\partial \widetilde{P}_3^2/\partial \widetilde{x}_1\right) + q_i\widetilde{P}_3\left(\left(\widetilde{P}_3^S\right)^2 - \widetilde{P}_3^2\right)$ ($i=1,2$) originate from the flexoelectric (term $\sim f_i^Q$) and electrostriction (term $\sim q_i$) couplings. Expressions for $f_i^Q$ and $q_i$ are also listed in the end of the **Table 1**. Coefficients $f_i^Q$ are directly proportional to the product of electrostriction and flexoelectric coupling coefficients and inversely proportional to the elastic compliances combination.

In the **Table 1** all terms proportional to the second powers of the flexoelectric coupling coefficients and their derivatives in $f_i^Q$ were omitted, i.e. neglect the terms proportional to $\widetilde{F}_{ij}^2$, $\widetilde{F}_{kl}F_{ij}$ and $\widetilde{F}_{ij}^2$. In particular, the polarization variation $\delta \widetilde{P}_3(\widetilde{x}_1)$ as proportional to the second powers of the flexoelectric coupling coefficient was omitted. Thus for all types of the walls $\widetilde{P}_3^S \approx \widetilde{P}_3^S \tanh(\widetilde{x}_1/L_c)$, where the function $\tanh(\widetilde{x}_1/L_c)$ should be used for the materials with the second order ferroelectric phase transition, like BiFeO$_3$. For the materials with the first order ferroelectric phase transition it should be substituted with $\dfrac{\sinh(\widetilde{x}_1/L_c)}{\sqrt{\cosh^2(\widetilde{x}_1/L_c)+c}}$, where the constant $c \leq 0.5$ [33].

Note that expression for $\widetilde{P}_2$ gives an overestimated value that does not account for gradient effects. A more rigorous expression including this effect is derived by Yudin et al. [34] for the case of tetragonal BaTiO$_3$. Linear variations $\delta \widetilde{P}_2(\widetilde{x}_1)$ are absent for 109- and 71-degree uncharged domain walls, since $f_2^Q(\alpha) \sim \sin(4\alpha)$ and $q_2(\alpha) \sim \sin(4\alpha)$ are zero for the corresponding angles $\alpha = 0, \pi$ and $\alpha = -\pi/4, 3\pi/4$.

The contributions $\widetilde{P}_1(\widetilde{x}_1) \sim \dfrac{\varepsilon_0 \varepsilon_b q_1 \widetilde{P}_3}{1+2\beta\varepsilon_0\varepsilon_b}\left(\left(\widetilde{P}_3^S\right)^2 - \widetilde{P}_3^2\right)$ and $\widetilde{P}_2(\widetilde{x}_1) \sim \dfrac{q_2}{2\beta}\widetilde{P}_3\left(\left(\widetilde{P}_3^S\right)^2 - \widetilde{P}_3^2\right)$ originate from electrostriction coupling term proportional to $\widetilde{P}_3\left(\left(\widetilde{P}_3^S\right)^2 - \widetilde{P}_3^2\right) \approx \dfrac{L_c \widetilde{P}_3^S}{2}\dfrac{\partial \widetilde{P}_3^2}{\partial \widetilde{x}_1}$ across 180-degree walls in rhombohedral phase. They are not negligible for arbitrary angles $\alpha$, however they vanish for several definite angles. For instance $q_1 = 0$ for the angles $\alpha = m\pi/3$ corresponding to the case $\widetilde{Q}_{15} = \widetilde{F}_{15} = 0$; $q_2 = 0$ for the angles $\alpha = m\pi/3 + \pi/2$ corresponding to the case $\widetilde{s}_{14} = \widetilde{Q}_{14} = 0$ ($m$ is integer).

These results imply that the *nominally* uncharged 180-degree wall cannot be regarded completely uncharged for arbitrary angle $\alpha$, since the electrostriction-related term $\dfrac{\varepsilon_0 \varepsilon_b q_1 \widetilde{P}_3}{1+2\beta\varepsilon_0\varepsilon_b}\left(\left(\widetilde{P}_3^S\right)^2 - \widetilde{P}_3^2\right)$ induces the Neel-type component $\widetilde{P}_1$, corresponding bound charge and electric field. The latter can be rather small, but (as will be shown below) enough to induce carrier accumulation.

Note that Bloch-type walls were calculated previously in rhombohedral BaTiO$_3$ at low temperature [35] and tetragonal PbTiO$_3$ [36] without consideration of the flexoelectric coupling. Therefore it is assumed that a nonzero component $\widetilde{P}_2 = P_b \cosh^{-2}(\widetilde{x}_1/L_c)$ is possible in rhombohedral phase of BiFeO$_3$. The "even-type" solution $P_b \cosh^{-2}(\widetilde{x}_1/L_c)$ is unrelated with flexoelectric coupling and is symmetric with respect to $\widetilde{x}_1$. Since no compact analytical expression for $P_b$ can be derived, its existence was studied numerically. Numerical simulations showed that the even-type Bloch solution can appear at angles $\alpha = m\pi/3 + \pi/2$ ($m$ is integer), corresponding to the absence of the seeding, $f_2^Q(\partial \widetilde{P}_3^2/\partial \widetilde{x}_1) + q_2 \widetilde{P}_3 ((\widetilde{P}_3^S)^2 - \widetilde{P}_3^2)$, since $f_2^Q = q_2 = 0$, $\widetilde{s}_{14} = 0$ and $\widetilde{Q}_{14} = 0$ at these angles simultaneously. However, the even-type seeding is required for $P_b \cosh^{-2}(\widetilde{x}_1/L_c)$ origin, while the amplitude $P_b$ is the seeding-independent. So, the appearance of $P_b \cosh^{-2}(\widetilde{x}_1/L_c)$, as the typical manifestation of the spontaneous symmetry breaking across the wall, should be essentially energetically preferable in order to be dominant. Meanwhile the odd-type solution is induced by the "seeding" $f_2^Q(\partial \widetilde{P}_3^2/\partial \widetilde{x}_1) + q_2 \widetilde{P}_3 ((\widetilde{P}_3^S)^2 - \widetilde{P}_3^2)$ proportional to $f_2^Q$ and $q_2$, and is energetically preferable at all angles $\alpha \neq m\pi/3 + \pi/2$. Estimations of the corresponding wall energies proved that the even-type solution gives no more than several % energy gain in comparison with the odd-type solution at angles $\alpha = m\pi/3 + \pi/2$. Moreover, the even-type Bloch solution energy and features appeared strongly dependent on the gradient tensor $g_{ij}$ values and anisotropy. Since exact values and anisotropy of the gradient tensor is unknown for BiFeO$_3$, below $P_b = 0$ is substituted, and the question of the odd-type (i.e. $\widetilde{P}_2 \sim f_2^Q(\partial \widetilde{P}_3^2/\partial \widetilde{x}_1) + q_2 \widetilde{P}_3 ((\widetilde{P}_3^S)^2 - \widetilde{P}_3^2)$) and even-type (i.e. $\widetilde{P}_2 \sim P_b \cosh^{-2}(\widetilde{x}_1/L_c)$) Bloch solutions is left for future study.

The Neel odd-type component $\widetilde{P}_1(\widetilde{x}_1)$ is affected by the depolarization field $\widetilde{E}_1^d(\widetilde{x}_1) = -\partial\varphi/\partial\widetilde{x}_1$. Corresponding electrostatic potential $\varphi$ can be determined from the Poisson equation, $\varepsilon_0 \varepsilon_b \frac{\partial^2 \varphi}{\partial \widetilde{x}_i^2} = \frac{\partial P_i}{\partial \widetilde{x}_i} - e(p - n + N_d^+ - N_a^-)$, where $\varepsilon_0 = 8.85 \times 10^{-12}$ F/m is the universal dielectric constant, $\varepsilon_b$ is the background dielectric permittivity unrelated with the soft mode permittivity, absolute value of the electron charge $e=1.6\times10^{-19}$ C; $p(\varphi)$ and $n(\varphi)$ are electrons and holes density; $N_d^+$ and $N_a^-$ are the concentration of ionized acceptors and donors correspondingly. The bare (i.e. un-screened) depolarization field $\tilde{E}_1^d(\tilde{x}_1) \approx -\left(\tilde{P}_1(\tilde{x}_1) - \overline{\tilde{P}_1(\tilde{x}_1)}\right)/\varepsilon_0\varepsilon_b$ is caused by the inhomogeneity of $\tilde{P}_1(\tilde{x}_1)$ that originated from the flexoelectric effect and electrostriction, e.g. $\delta\tilde{P}_1(\tilde{x}_1) \sim \frac{\varepsilon_0\varepsilon_b f_1^Q}{1+2\beta\varepsilon_0\varepsilon_b}\frac{\partial \tilde{P}_3^2}{\partial \tilde{x}_1} + \frac{\varepsilon_0\varepsilon_b q_1 \tilde{P}_3}{1+2\beta\varepsilon_0\varepsilon_b}\left(\left(\tilde{P}_3^S\right)^2 - \tilde{P}_3^2\right)$ for 180-degree walls. Here $\overline{\tilde{P}_1(\tilde{x}_1)}$ is the spatial averaging; $\varepsilon_b$ is the background dielectric constant [37] unrelated with soft mode; $\varepsilon_0 = 8.85\times 10^{-12}$ F/m. Depolarization field corresponds to the electric potential variation $\delta\varphi(\tilde{x}_1)$:

$$\delta\varphi(\tilde{x}_1) = -\int_{-\infty}^{\tilde{x}_1} dx \tilde{E}_1^d(x) \approx \frac{\left(f_1^Q + q_1 L_c \tilde{P}_3^S/2\right)\left(P_3^2(\tilde{x}_1) - \left(\tilde{P}_3^S\right)^2\right)}{1+2\beta\varepsilon_0\varepsilon_b}. \qquad (3)$$

Constant $\beta$ is defined in the **Table 1**. Since for many ferroelectrics $2\beta\varepsilon_0\varepsilon_b \ll 1$, the potential variation $\delta\varphi(\tilde{x}_1) \approx \left(f_1^Q + q_1 L_c \tilde{P}_3^S/2\right)\left(P_3^2(\tilde{x}_1) - \left(\tilde{P}_3^S\right)^2\right)$ appeared almost independent on the background permittivity $\varepsilon_b$. The potential variation in turn changes the electrochemical potential in the vicinity of domain wall.

The coupling between the inhomogeneous strain and band structure is given by deformation potential [38, 39, 40, 41]. The strain-induced conduction and valence band edge shift caused by the domain wall is proportional to the strain variation $\delta u_{ij}(\mathbf{r}) = u_{ij}(\mathbf{r}) - u_{ij}^S$, where $u_{ij}^S$ is the spontaneous strain. Assuming the linear dependence [42]:

$$E_C(\delta u_{ij}(\mathbf{r})) = E_{C0} + \Xi_{ij}^C \delta u_{ij}(\mathbf{r}), \qquad E_V(\delta u_{ij}(\mathbf{r})) = E_{V0} + \Xi_{ij}^V \delta u_{ij}(\mathbf{r}). \qquad (4)$$

where $E_C$ and $E_V$ are the energetic position of the bottom of conduction band and the top of the valence band respectively, $\Xi_{ij}^{C,V}$ is a tensor deformation potential of electrons in the conduction (*C*) and valence bands (*V*). Values $E_{C0} = E_C(u_{ij}^S)$ and $E_{V0} = E_V(u_{ij}^S)$ already includes the spontaneous strain $u_{ij}^S$ existing far from the domain wall.

The symmetry of deformation potential tensors $\Xi_{ij}^{C,V}$ in the $\Gamma$-point is determined by the crystal spatial symmetry [42]. In particular, the components $\Xi_{11}^{C,V} = \Xi_{22}^{C,V} = \Xi_{33}^{C,V}$ and $\Xi_{12}^{C,V} = \Xi_{23}^{C,V} = \Xi_{13}^{C,V}$ can be nonzero in the rhombohedral phase of bulk ferroelectric $BiFeO_3$. For the case of ferroelectric in tetragonal phase (strained films of $BiFeO_3$, bulk $PbZr_{0.2}Ti_{0.8}O_3$ and $BaTiO_3$ at room temperature) the nonzero components are $\Xi_{11}^{C,V} = \Xi_{22}^{C,V}$ and $\Xi_{33}^{C,V}$ in the crystallographic frame, and the tensor is diagonal $\Xi_{ij}^{C,V} \equiv \Xi_{11}^{C,V} \delta_{ij}$.

The band edge shift Eqs.(4), induced by the inhomogeneous strain Eqs.(1) via the deformation potential, and electric potential variation Eq.(3), induced by the flexoelectric coupling in the vicinity of uncharged domain wall, modulates the densities of free electrons $n(\tilde{x}_1)$ and holes $p(\tilde{x}_1)$ accumulated by the domain wall. The effect can be estimated in the Boltzmann approximation as [43]:

$$n(\tilde{x}_1) = \int_0^\infty d\varepsilon \cdot g_n(\varepsilon) f\big(\varepsilon + E_C(u_{ij}) - E_F(u_{ij}) - e\varphi\big) \approx n_0 \exp\left(\frac{\Delta E_n(\tilde{x}_1)}{k_B T}\right), \quad (5a)$$

$$p(\tilde{x}_1) = \int_0^\infty d\varepsilon \cdot g_p(\varepsilon) f\big(-\varepsilon - E_V(u_{ij}) + E_F(u_{ij}) + e\varphi\big) \approx p_0 \exp\left(\frac{\Delta E_p(\tilde{x}_1)}{k_B T}\right). \quad (5b)$$

Here $k_B = 1.3807 \times 10^{-23}$ J/K, $T$ is the absolute temperature, $f(x) = \{1 + \exp(x/k_B T)\}^{-1}$ is the Fermi-Dirac distribution function, electron charge $e = 1.6 \times 10^{-19}$ C; local band bending for electrons ($\Delta E_n$) and holes ($\Delta E_p$) are introduced in Eqs.(5) as

$$\Delta E_n(\tilde{x}_1) = -\Xi_{ij}^n \delta u_{ij}(\tilde{x}_1) + e\delta\varphi(\tilde{x}_1), \qquad \Delta E_p(\tilde{x}_1) = \Xi_{ij}^p \delta u_{ij}(\tilde{x}_1) - e\delta\varphi(\tilde{x}_1). \quad (6a)$$

"Effective" deformation potentials are introduced as:

$$\Xi_{ij}^n = \Xi_{ij}^C - \xi_{ij}^F, \qquad \Xi_{ij}^p = \Xi_{ij}^V - \xi_{ij}^F. \quad (6b)$$

The equilibrium densities of holes and electrons are

$$n_0 = \int_0^\infty d\varepsilon \cdot g_n(\varepsilon) \exp\left(-\frac{\varepsilon + E_C(u_{ij}^S) - E_F(u_{ij}^S)}{k_B T}\right), \quad (6c)$$

$$p_0 = \int_0^\infty d\varepsilon \cdot g_p(\varepsilon) \exp\left(\frac{\varepsilon + E_V(u_{ij}^S) - E_F(u_{ij}^S)}{k_B T}\right). \quad (6d)$$

Fermi level position $E_F(u_{ij}^S)$ should be determined self-consistently from the electro-neutrality condition $p_0 - n_0 + N_d^+ - N_a^- = 0$ valid in the single-domain region of ferroelectric, where $\delta\varphi = 0$ and $\delta u_{ij} = 0$. The concentrations of almost immobile ionized acceptors and donors are $N_d^+ \approx N_{d0}^+ \exp\left(\frac{E_{d0} - E_F(u_{ij}^S)}{k_B T}\right)$ and $N_a^- \approx N_{a0}^- \exp\left(\frac{E_F(u_{ij}^S) - E_{a0}}{k_B T}\right)$ correspondingly. For mobile species Vegard strains should be included in expressions for $N_d^+$ and $N_a^-$ [43].

When deriving Eqs.(6), included in the analysis is the fact that Fermi level position can be strain-dependent in the semiconductor with strongly prevailing improper conductivity [44], e.g. one can assume the power expansion and cut it on the linear term, $E_F(u_{ij}) = E_{F0}(u_{ij}^S) + \xi_{ij}^F \delta u_{ij}$. For the case the dependence $E_F(u_{ij})$ can originate from the fact that donor and acceptor level positions $E_d$ and $E_a$ can be strain-dependent [44], e.g. $E_d = E_{d0}(u_{ij}^S) + \xi_{ij}^d \delta u_{ij}$, and the dependence can lead to the strain-dependent Fermi level in the local density approximation. For instance, in the model case of the non-degenerated simple band structure and effective mass approximation validity, the shallow donor level $E_d$ and conductive band edge $E_C$ can be shifted as a whole with the strain increase due to deformation potential effect [44], i.e. $\xi_{ij}^d \approx \Xi_{ij}^C \approx \xi_{ij}^F$. In the latter case the difference $E_F(u_{ij}) - E_C(u_{ij}) \approx E_{F0} - E_{C0}$ is almost strain-independent leading to the almost strain-independent $\Delta E_n \approx e\delta\varphi$ in Eqs.(6a), for purely improper electronic conductivity. The situation $\Delta E_p \approx -e\delta\varphi$ is also not excluded in general. The exact strain dependence of $\Delta E_{p,n}$ is unknown for BiFeO$_3$, where the band structure is extremely complex and degenerated, and improper conductivity is typically of mixed p-type [45]. Hence, below we explore both cases in Eqs.(6); the case $\Xi_{ij}^p = 0$ in comparison with $\Xi_{ij}^p \neq 0$, since estimations for the band gap derivative $\partial E_g / \partial u_{ij} \sim 20$ eV are available [46].

Equations (5)-(6) show that there are two contributions into the bands bending and carrier density variation across the uncharged domain wall: the first term $\mp \Xi_{ij}^{n,p} \delta u_{ij}(\tilde{x}_1)$ is proportional to the deformation potential, the second term $\pm e\delta\varphi(\tilde{x}_1)$ originates from the flexoelectric effect. Using perturbation approach on the flexoelectric coupling value, it is possible to simplify Eqs. (5) further in order to estimate and compare both contributions. The simplifications are possible using approximate expression for electric potential variation $\delta\varphi(\tilde{x}_1) \approx (f_1^Q + q_1 L_c \tilde{P}_3^S / 2)(P_3^2(\tilde{x}_1) - (\tilde{P}_3^S)^2)$, diagonal deformation potential tensor, $\Xi_{ij}^{n,p} \equiv \Xi_{11}^{n,p} \delta_{ij}$ and evident form of elastic strains Eqs. (1), that gives $\Xi_{ij}^{n,p} \delta u_{ij}(\tilde{x}_1) \equiv \Xi_{11}^{n,p} \delta u_{11}(\tilde{x}_1)$. If $P_3(\tilde{x}_1)$ contribution dominates for the materials with the second order transition into ferroelectric phase, expressions from the **Table 1** allows us to put $((\tilde{P}_3^S)^2 - P_3^2(\tilde{x}_1)) = (\tilde{P}_3^S)^2 \cosh^{-2}(\tilde{x}_1/L_c)$ and $\partial P_3^2(\tilde{x}_1)/\partial \tilde{x}_1 = 2((\tilde{P}_3^S)^2/L_c)\sinh(\tilde{x}_1/L_c)\cosh^{-3}(\tilde{x}_1/L_c)$ for rough estimations in Eqs.(5). After elementary transformations of Eqs.(5)-(6) an estimation for the carrier accumulation by 180-degree domain wall is derived:

$$n(\tilde{x}_1) \approx n_0 \exp\left(\frac{\Xi_{11}^n(\vartheta_{13} + (2/L_c)(\Psi_{13} + \vartheta_{11} f_1^Q \varepsilon_0 \varepsilon_b)\tanh(\tilde{x}_1/L_c)) - e(f_1^Q + q_1 L_c \tilde{P}_3^S/2)}{k_B T \cdot \cosh^2(\tilde{x}_1/L_c)}(\tilde{P}_3^S)^2\right), \quad (7a)$$

$$p(\widetilde{x}_1) \approx p_0 \exp\left(\frac{-\Xi_{11}^p\left(\vartheta_{13} + (2/L_c)\left(\Psi_{13} + \vartheta_{11}f_1^Q\varepsilon_0\varepsilon_b\right)\tanh(\widetilde{x}_1/L_c)\right) + e\left(f_1^Q + q_1 L_c \widetilde{P}_3^S/2\right)}{k_B T \cdot \cosh^2(\widetilde{x}_1/L_c)}\left(\widetilde{P}_3^S\right)^2\right). \quad (7b)$$

Note, that Eqs.(7) are linear approximation on the flexoelectric coupling, since here all terms proportional to the second powers of the flexoelectric coupling coefficients and their derivatives are omitted. It is seen from Eqs.(7) that the maximal (or minimal) carrier density corresponds to the vicinity of wall plane $\widetilde{x}_1 = 0$. Simple expressions are valid:

$$k_B T \ln\left(\frac{n(0)}{n_0}\right) \sim \left(\Xi_{11}^n \vartheta_{13} - e\left(f_1^Q + q_1 L_c \frac{\widetilde{P}_3^S}{2}\right)\right)\left(\widetilde{P}_3^S\right)^2 \quad (8a)$$

$$k_B T \ln\left(\frac{p(0)}{p_0}\right) \sim \left(-\Xi_{11}^p \vartheta_{13} + e\left(f_1^Q + q_1 L_c \frac{\widetilde{P}_3^S}{2}\right)\right)\left(\widetilde{P}_3^S\right)^2. \quad (8b)$$

Strong dependence of the coefficients $\vartheta_{ij}(\alpha)$, $f_1^Q(\alpha)$ and $q_1(\alpha)$ of the wall rotation angle $\alpha$ explores the anisotropic nature of the carrier accumulation by uncharged 180-degree domain walls, where $\alpha$ can be arbitrary.

More rigorous Eqs (5) allow one to estimate quantitatively the relative contributions of the deformation potential, electrostriction and flexoelectric coupling in the static conductivity of uncharged domain walls. Electrostriction coefficients and elastic compliances are relatively well-known for typical ferroelectrics (see **Table S2**). Numerical values of the deformation potential tensor components are poorly known for ferroelectrics. Below the estimation $\left|\Xi_{ij}^{C,V}\right| \sim (5 - 20)$ eV is used, consistent with experimental result for BiFeO$_3$ [46] and *ab-initio* calculations for SrTiO$_3$ [47].

Flexoelectric coefficient values $F_{ij}$ can be estimated as $\sim 10^{-11}$C$^{-1}$m$^3$ from experiment for SrTiO$_3$ [48], and vary in the range $(1 - 100) \times 10^{-11}$C$^{-1}$m$^3$ for BaTiO$_3$ [49], $F_{ij} \sim 300 \times 10^{-11}$C$^{-1}$m$^3$ for PbZr$_{0.2}$Ti$_{0.8}$O$_3$ [50]. Kogan microscopic model [51] gives $F_{ij} \sim 10^{-11}$C$^{-1}$m$^3$ for all perovskites with lattice constant 0.4 nm. For BiFeO$_3$ with parent m3m symmetry nonzero components of the flexoelectric tensor are $F_{11}$, $F_{12}$, $F_{44}$. Using the estimations for $\Xi_{ij}^{n,p}$, $F_{ij}$ and parameters from the **Table S2**, it is seen that the contributions of the flexoelectric coupling and deformation potential in the uncharged domain wall conductivity in BiFeO$_3$ are comparable and can lead to one to two orders of magnitude increase of the wall static conductivity.

### 3. Results for flat walls

Angular dependences of local band bending and holes density accumulated by 180-degree domain walls in p-type BiFeO$_3$ are shown in the **Figs 3** for different values of flexoelectric coupling coefficients $F_{ij}$ and deformation potential $\Xi_{ij}^p$. For the case $\Xi_{ij}^p = 0$ and $F_{ij} = 0$ (dotted curves) strong anisotropy originates from the angular dependence of the electrostriction coefficients (see e.g. expressions for $q_{1,2}$ in the **Table 1**) and corresponding band bending angular dependence is quasi-harmonic, $\Delta E_p(\tilde{x}_1 = 0) \sim E_0 \sin(3\alpha)$. One can see from the **Fig. 3a** that flexoelectric coupling shifts the angular dependence as $\Delta E_p(\tilde{x}_1 = 0) \sim E_0 \sin(3\alpha) + E_1$, where the shift amplitude $E_1$ is proportional to $F_{ij}$ combinations. Thus band bending and holes density are minimal at the angles $\alpha = 3\pi/6, 7\pi/6, 11\pi/6$ and maximal at the angles $\alpha = \pi/6, 5\pi/6, 9\pi/6$. Bulk level $\Delta E_p = 0$ and $p = p_0$ corresponds to the angles $\alpha = 0, \pi/3, 2\pi/3$ at $F_{ij} = 0$ and weakly shifts $F_{ij}$ increase. Modulation depth of the carriers accumulation/depletion is about 3-4 orders depending on $F_{ij}$ values. In particular the holes density at the wall can increase by a factor of 50 in comparison with a bulk value at $\Xi_{ij}^p = 0$ and $F_{ij} = 0$; by a factor of 100 for realistic values of the flexoelectric coefficients $F_{11} = -1.38 \times 10^{-11} C^{-1} m^3$, $F_{12} = 0.67 \times 10^{-11} C^{-1} m^3$, $F_{44} = 0.85 \times 10^{-11} C^{-1} m^3$ corresponding to SrTiO3 [48] and effective deformation potential $\Xi_{ij}^p = 21$ eV estimated for BiFeO3 from experiment [46]. The accumulation effect appeared relatively insensitive to the deformation potential value (compare dotted black curves for $\Xi_{ij}^p = 0$ with red curves for $\Xi_{ij}^p = 21$ eV). Polar plot **Fig. 3c** demonstrates 6-lobe structure, where 3 longer lobes (solid curves) correspond to positive band bending $\Delta E_p$ and 3 smaller lobes (dashed curves) correspond to negative $\Delta E_p$. The difference increases between sizes of "positive" and "negative" lobes increases with the amplitude of the flexoelectric coupling increase. Polar plot **Fig. 3d** demonstrates 3-lobe structure, where each lobe corresponds to the holes accumulation ($p >> p_0$) by the wall.

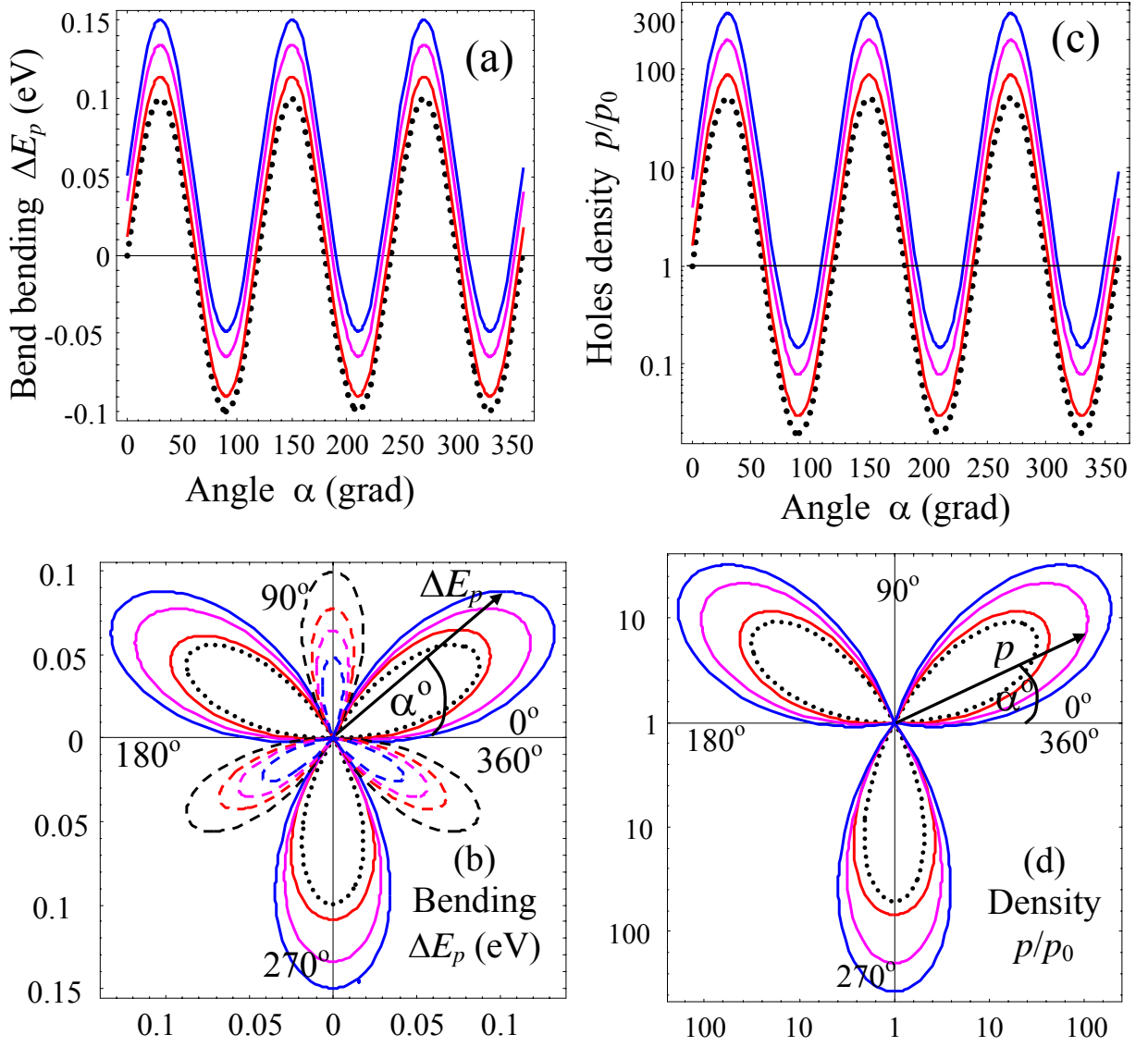

**Figure 3.** Anisotropic local band bending $\Delta E_p(0)$ (a,b) and holes density $p(0)/p_0$ (c,d) angular dependences caused by nominally uncharged 180-degree domain walls in rhombohedral multiferroic BiFeO$_3$. **Black** dotted curves in the angular dependences (a,c) and polar plots (b,d) are calculated at the wall plane $\tilde{x}_1 = 0$ without deformation potential $\Xi_{ij}^p = 0$ and flexoelectric effect $F_{ij} = 0$. Solid curves are calculated for different flexoelectric coefficients: $F_{11} = -1.38 \times 10^{-11}$C$^{-1}$m$^3$, $F_{12} = 0.67 \times 10^{-11}$C$^{-1}$m$^3$, $F_{44} = 0.85 \times 10^{-11}$C$^{-1}$m$^3$ and $\Xi_{ij}^p = 0$ (**red** solid curves); $2F_{ij}$ and $\Xi_{ij}^p = 21$ eV (**magenta** solid curves); $3F_{ij}$ and $\Xi_{ij}^p = 21$ eV (**blue** solid curves). Dashed curves in the polar plot (c) correspond to the negative $\Delta E_p(\tilde{x}_1)$. Other parameters are listed in the **Table S2** at room temperature 293°K.

Polarization components, strain, potential, band bending and holes density $\tilde{x}_1$-profiles across the 180-degree domain wall were calculated for $\Xi_{ij}^p = 0$ and $F_{ij} = 0$ (as shown in **Figs.4**) and for $\Xi_{ij}^p = 21$ eV and $F_{11} = -4.14 \times 10^{-11}$C$^{-1}$m$^3$, $F_{12} = 2.01 \times 10^{-11}$C$^{-1}$m$^3$, $F_{44} = 2.55 \times 10^{-11}$C$^{-1}$m$^3$ (as shown in **Figs.5**). The values $F_{ij}$ are 3 times higher than the ones for SrTiO$_3$ [48]. Wall rotation angle α varies in the range 0 – 30°; the angular range corresponds to the ½ of the first lobe shown in **Figs.3c,d.** It is seen from the **Figs.4a** and **5a** that polarization components $\tilde{P}_1$ and $\tilde{P}_2$ are strongly dependent on the angle α and weakly dependent on the flexoelectric coupling strength. Spatial distributions of $\tilde{P}_1$ and $\tilde{P}_2$ are anty-symmetric with respect to the wall plane $\tilde{x}_1 = 0$. For all cases $\tilde{P}_2$ value is at least 3 times higher than $\tilde{P}_1$ one, since $\tilde{P}_2$ is not suppressed by the depolarization field $\tilde{E}_1^d(\tilde{x}_1)$. It is seen from the **Figs.4b and 5b** that polarization component $\tilde{P}_3$ is almost independent on the angle α as anticipated from the **Table 1**. The strain $\delta u_{11}(\tilde{x}_1)$ is dependent on the flexoelectric coupling, but its maximal value is almost independent on the angle α [compare **Figs.4c** and **5c**]. Dependences of the strain $\delta u_{11}(\tilde{x}_1)$ and potential $\delta\varphi(\tilde{x}_1)$ on $\tilde{x}_1$ are bell-shaped and symmetric with respect to the wall plane $\tilde{x}_1 = 0$ [see **Figs.4c,d** and **5c,d**]. Contribution of the potential variation $-e\delta\varphi$ into the band bending $\Delta E_p = \Xi_{ij}^p \delta u_{ij} - e\delta\varphi$ typically dominates over the strain contribution $\Xi_{11}^p \delta\tilde{u}_{11}$ [compare values and profiles shape in **Figs.4c,d,e and 5c,d,e**]. Therefore the holes accumulation by the 180-degee wall is governed by the potential variation. Flexoelectric coupling increase enhances the strain, potential variation and band bending maximal values, but the shape of their $\tilde{x}_1$-profiles remains almost the same (compare scales in **Figs.4f and 5f**).

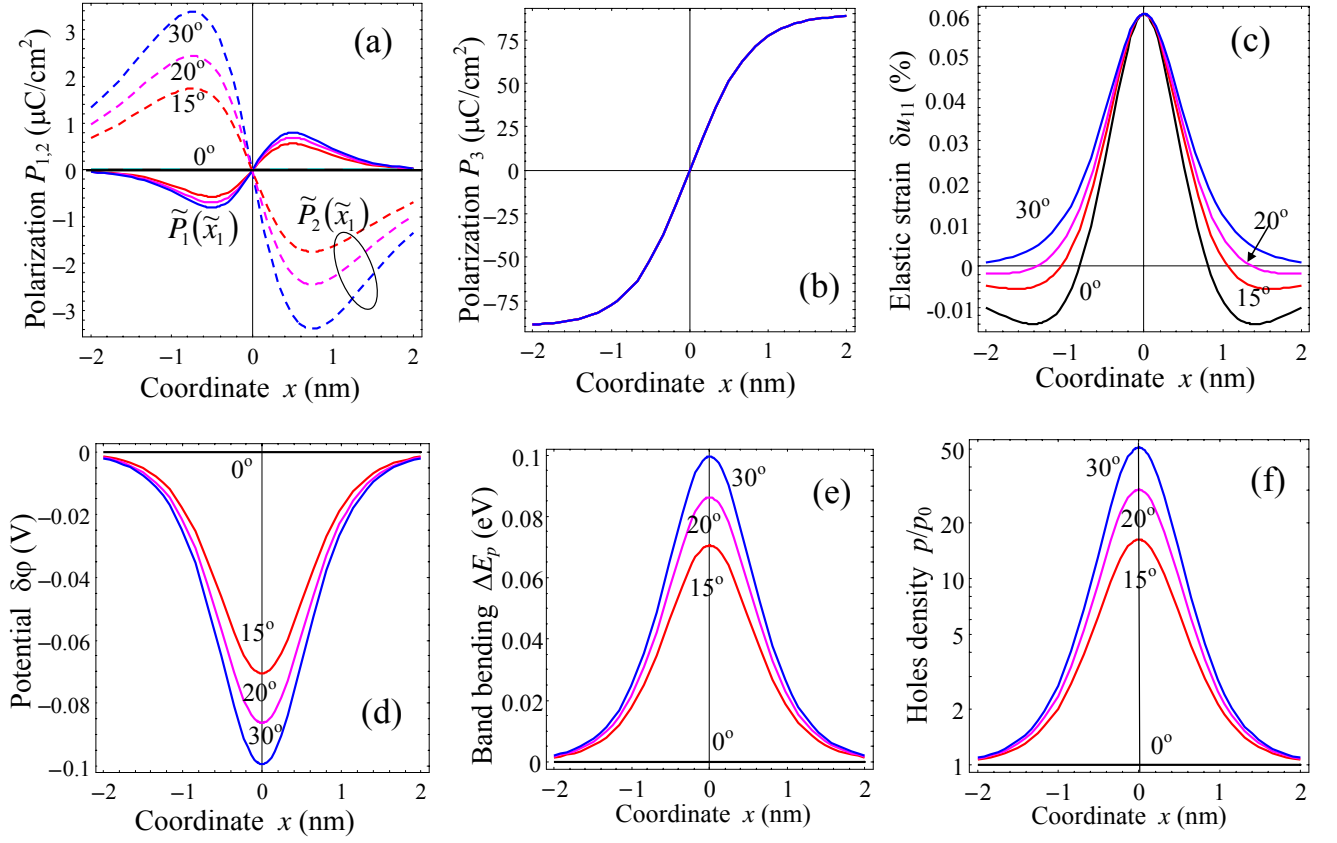

**Figure 4.** Profiles of polarization components $\tilde{P}_1(\tilde{x}_1)$ (a, solid curves), $\tilde{P}_2(\tilde{x}_1)$ (a, dashed curves) and $\tilde{P}_3(\tilde{x}_1)$ (b), elastic strain $\delta u_{11}(\tilde{x}_1)$ (c), potential variation $\delta\varphi(\tilde{x}_1)$ (d), local band bending $\Delta E_p(\tilde{x}_1)$ (e) and carrier density $p(\tilde{x}_1)/p_0$ (f) calculated across nominally uncharged 180-degree domain walls in rhombohedral BiFeO$_3$ without deformation potential $\Xi_{ij}^p = 0$ and flexoelectric effect $F_{ij} = 0$. Rotation angle $\alpha = 0$ (**black** curves), $\alpha = 15°$ (**red** curves), $\alpha = 20°$ (**magenta** curves), $\alpha = 30°$ (**blue** curves). Material parameters are listed in the **Table S2** at room temperature 293°K.

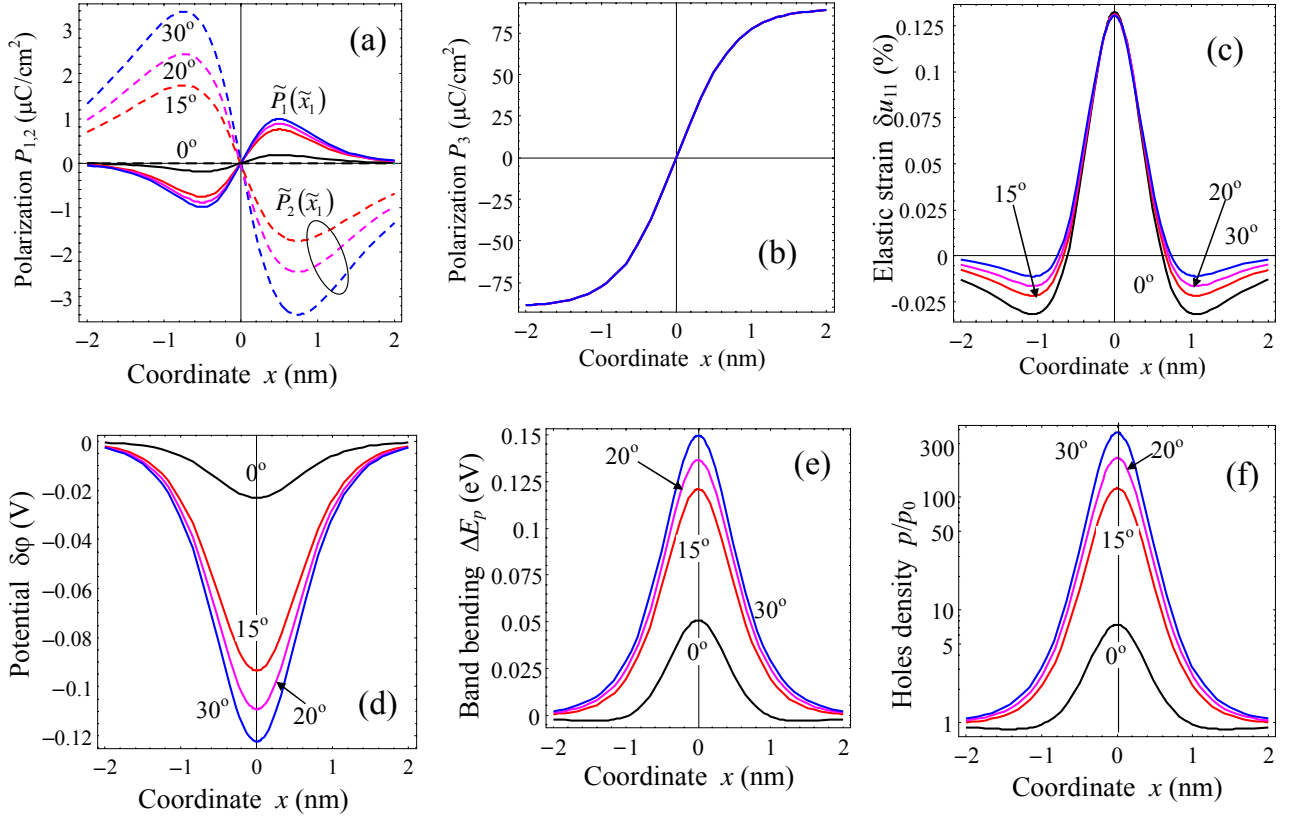

**Figure 5.** Profiles of polarization components $\tilde{P}_1(\tilde{x}_1)$ (a) and $\tilde{P}_3(\tilde{x}_1)$ (b), the elastic strain $\delta u_{11}(\tilde{x}_1)$ (c), potential variation $\delta\varphi(\tilde{x}_1)$ (d), local band bending $\Delta E_p(\tilde{x}_1)$ (e) and carrier density $p(\tilde{x}_1)/p_0$ (f) calculated across nominally uncharged 180-degree domain walls in rhombohedral BiFeO$_3$ allowing for flexoelectric coupling coefficients $F_{11}= -4.14\times10^{-11}$C$^{-1}$m$^3$, $F_{12}=2.01\times10^{-11}$C$^{-1}$m$^3$, $F_{44}=2.55\times10^{-11}$C$^{-1}$m$^3$ and deformation potential $\Xi_{ij}^p = 21$eV. Rotation angle $\alpha =0$ (**black** curves), $\alpha = 15^\circ$ (**red** curves), $\alpha = 20^\circ$ (**magenta** curves), $\alpha = 30^\circ$ (**blue** curves). Material parameters are listed in the **Table S2** at room temperature 293°K.

Detailed study of the carrier accumulation/depletion by 71 and 109-degree uncharged domain walls in BiFeO$_3$ was performed using Eqs.(1)-(6) and **Table S1b** in the similar way as was carried out for the 180-degree domain walls. Representative results are shown in **Figs.6-7**. In contrast to the equilibrium 180-degree uncharged domain walls, where the rotation angle $\alpha$ can be arbitrary, it is not the case of 109-degree uncharged domain walls, where $\alpha = 0$ (or $\pi$) and 71-degree domain walls, where $\alpha = -\pi/4$ (or $3\pi/4$) in equilibrium (see **Fig. 2a** with **2b,c**). In local equilibrium bright regions on the hole density maps and maximum on the corresponding $\tilde{x}_1$-profiles are located around the 109-degree weakly charged domain walls [**Fig. 6a**] and 71-degree weakly charged domain walls [**Fig. 7a**]. It is worth to underline that, in contrast to 180-degree walls, the flexoelectric coupling decreases the hole accumulation effect for 109- and 71-walls (compare solid and dashed curves in **Figs.6b** and **7b**). This appears because the potential variation $-e\delta\varphi(\tilde{x}_1)$ due to its negative sign across the walls (see **Figs.6c-d,** and **7c-d**) counteracts to the holes accumulation induced by the positive strain term $\Xi_{11}^p \delta u_{11}(\tilde{x}_1)$, since $\delta u_{11}(\tilde{x}_1) > 0$ (see **Figs.6e-f,** and **7e-f**). Resulting band bending, $\Delta E_p(\tilde{x}_1) = \Xi_{ij}^p \delta u_{ij}(\tilde{x}_1) - e\delta\varphi(\tilde{x}_1)$, becomes smaller than the value $\Xi_{11}^p \delta u_{11}(\tilde{x}_1)$ calculated without flexoelectric coupling, since $\delta\varphi(\tilde{x}_1) = 0$ without flexoelectric coupling (compare solid and dashed curves in **Figs.6d** and **7d**). The width of potential $\delta\varphi(\tilde{x}_1)$ and strain $\delta u_{11}(\tilde{x}_1)$ maxima is about 1 nm. Thus, the charging of the nominally uncharged walls originates from the flexoelectric coupling that induces polarization component $\tilde{P}_1(\tilde{x}_1) \sim f_1^Q \left(\partial \tilde{P}_3^2 / \partial \tilde{x}_1\right)$, which in turn induces corresponding depolarization field $\tilde{E}_1^d(\tilde{x}_1)$ and potential variation $\delta\varphi(\tilde{x}_1)$.

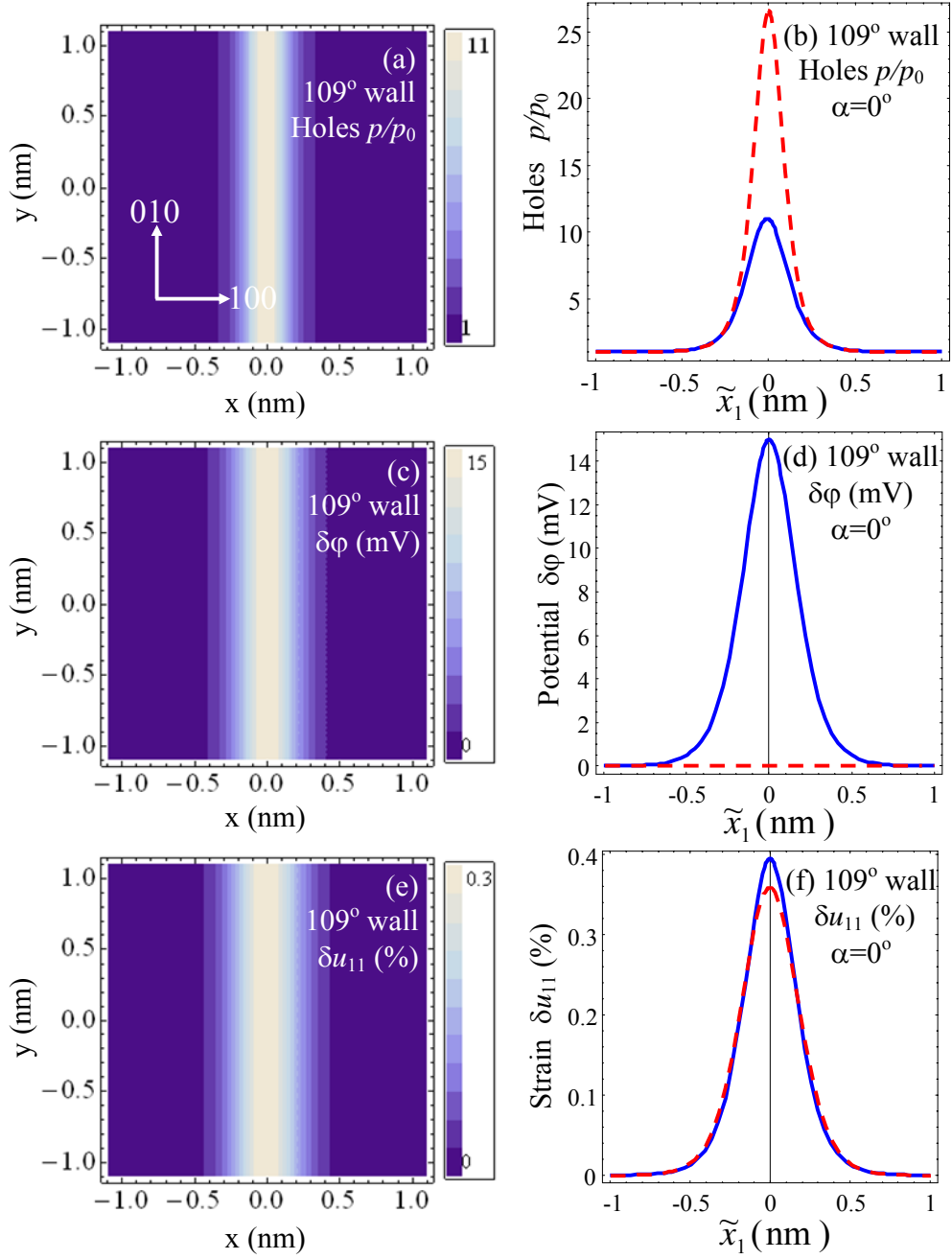

**Figure 6.** Contour maps and $\tilde{x}_1$-profiles of holes density $p(\tilde{x}_1)/p_0$ (a,b), potential variation $\delta\varphi(\tilde{x}_1)$ (c,d) and elastic strain $\delta u_{11}(\tilde{x}_1)$ (d,f) for 109-degree nominally uncharged domain walls. Maps (a,c,e) and **blue solid** profiles (b,d,f) are calculated for flexoelectric coupling $F_{11}=-1.38\times10^{-11}$C$^{-1}$m$^3$, $F_{12}=0.67\times10^{-11}$C$^{-1}$m$^3$, $F_{44}=0.85\times10^{-11}$C$^{-1}$m$^3$ and deformation potential $\Xi_{ij}^p=21$eV. **Red dashed** profiles (b,d,f) are calculated without flexoelectric coupling ($F_{ij}=0$).

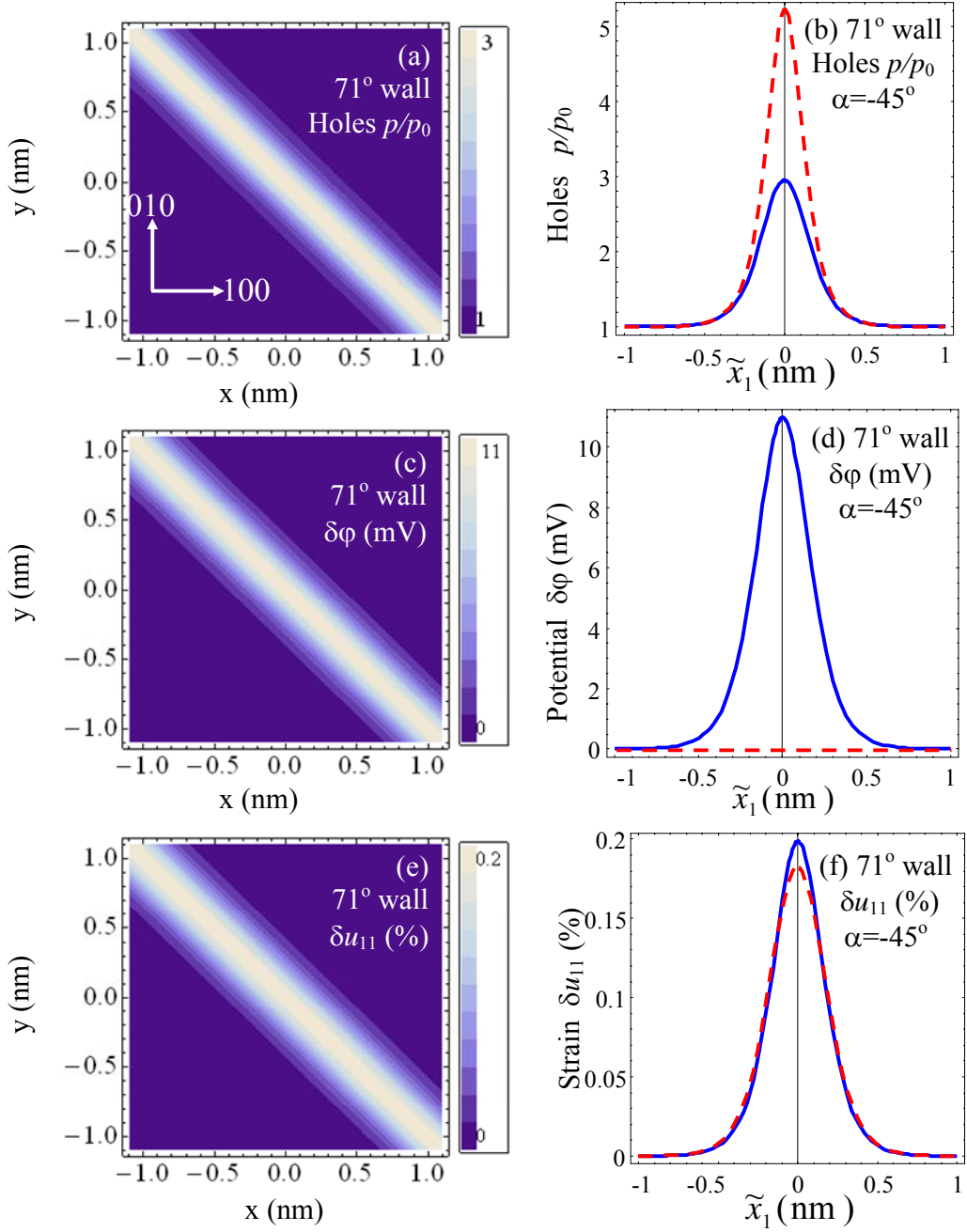

**Figure 7.** Contour maps and $\tilde{x}_1$-profiles of holes density $p(\tilde{x}_1)/p_0$ (a,b), potential variation $\delta\varphi(\tilde{x}_1)$ (c,d) and elastic strain $\delta u_{11}(\tilde{x}_1)$ (d,f) for 71-degree nominally uncharged domain walls. Maps (a,c,e) and **blue solid** profiles (b,d,f) are calculated for flexoelectric coupling $F_{11} = -1.38 \times 10^{-11} C^{-1} m^3$, $F_{12} = 0.67 \times 10^{-11} C^{-1} m^3$, $F_{44} = 0.85 \times 10^{-11} C^{-1} m^3$ and deformation potential $\Xi_{ij}^p = 21 eV$. **Red dashed** profiles (b,d,f) are calculated without flexoelectric coupling ($F_{ij}=0$).

Theoretical results presented in **Figs.3-7** are in good agreement with the first-principles study of ferroelectric domain walls in BiFeO$_3$ [20] that showed that band gap narrows on the value 0.2 eV, 0.1 eV and 0.05 eV at uncharged 180-, 109- and 71-degree domain walls correspondingly.

One of the simplest ways to verify experimentally our theoretical predictions of the 180-degree domain walls anisotropic conductivity is to study the conductivity on the cylindrical domain walls in BiFeO$_3$. Cylindrical domains can be readily created in BiFeO$_3$ and studied by the Scanning Probe Microscopy methods, such as c-AFM and PFM [52].

### 4. Cylindrical 180-degree domain wall

Evolved analytical theory can be extended for the case of cylindrical domain. For the case when domain radius $R$ is much higher than the correlation length $L_c$, using the method [53] for the nonlinear LGD-type equation solution in cylindrical geometry, relatively simple analytical expressions for the band bending were derived:

$$\Delta E_p(\alpha, \rho - R) \approx \Xi_{ij}^p \delta u_{ij}(\alpha, \rho - R) - e\delta\varphi(\alpha, \rho - R), \qquad (9)$$

where the polar radius $\rho = \sqrt{x^2 + y^2}$ and polar angle $\alpha = \arccos(x/\rho)$ are introduced [see **Fig. 8a**]. Eq.(8) predicts that the holes density, strain and potential spatial maps have the pronounced ring feature located at radii $|\rho - R| = L_c$ as shown in **Figs. 8b-d.** Three bright regions on the map in **Fig. 8b** correspond to the strong accumulation of holes; three dark regions are strongly depleted by holes. The ring's behaviour follows from the 3- and 6-lobes structure of the density and band bending polar plots shown in **Fig.3c,d**. Hereafter the static conductivity $\sigma(x, y)$ is regarded as proportional to the carriers densities as $\sigma(x, y) = e\mu_n n(x, y) + e\mu_p p(x, y)$, where mobilities $\mu_{n,p}$ are treated as constant. Thus the bright regions on the map in **Fig. 8b** correspond to the static conductivity enhancement; while the conductivity of the dark regions is much smaller than the bulk one. Also it is worth noting that elastic strain ring $\delta u_{11}(\rho)$ is almost homogeneous without bright and dark regions [**Fig. 8c**], while the potential ring has 3 dark and 3 bright regions, where positions are inverted with respect to the hole's density ring [**Fig. 8d**]. The ring's behaviour can be explained from the fact that potential variation contribution $-e\delta\varphi$ dominates over the strain contribution $\Xi_{11}^p \delta\tilde{u}_{11}$ into the band bending $\Delta E_p = \Xi_{ij}^p \delta u_{ij} - e\delta\varphi$ [see Eq.(4) and **Figs.5c,d,e**].

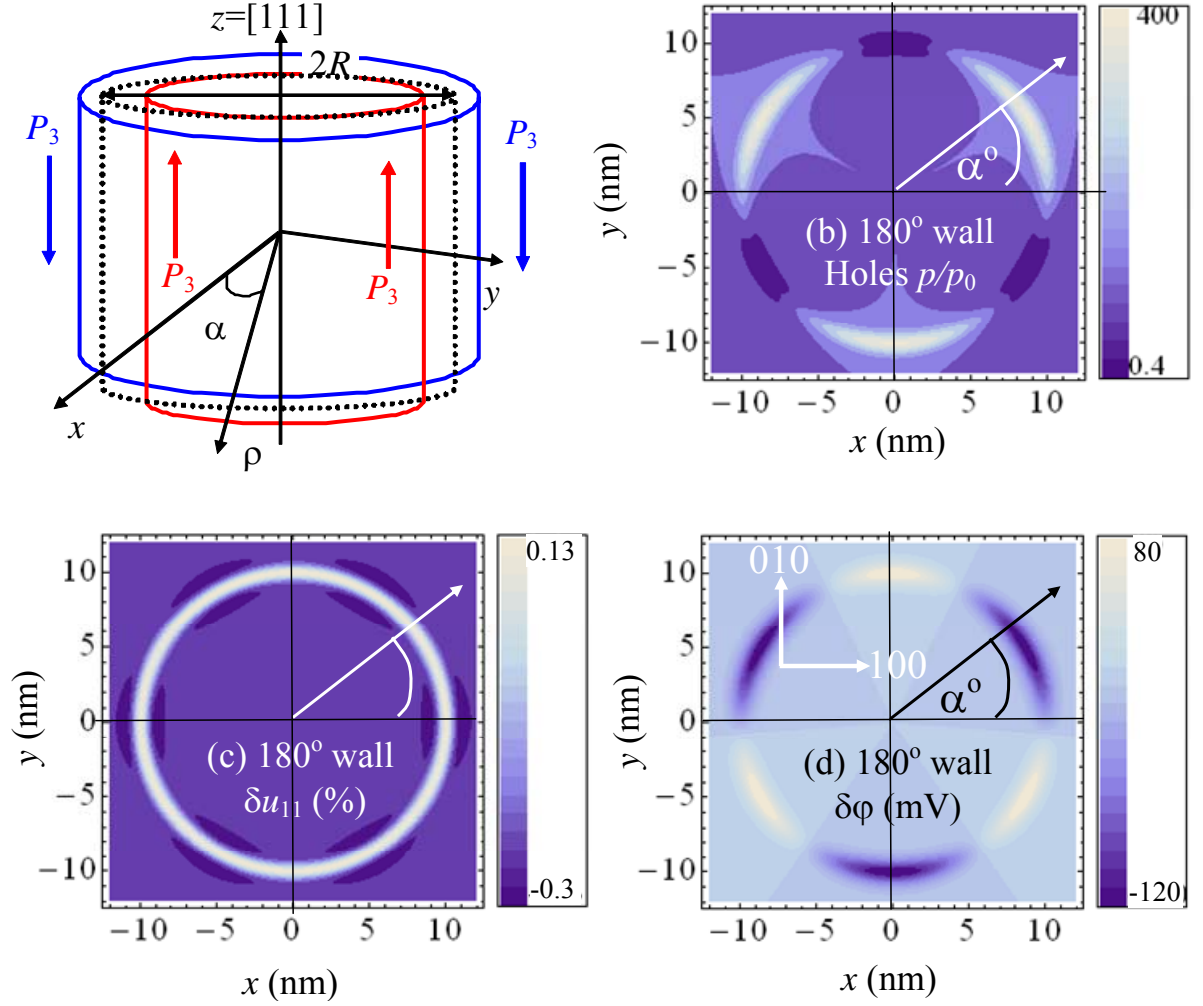

**Figure 8.** (a) Schematic of the cylindrical domain. Contour maps of the holes density $p(\rho)/p_0$ (b), elastic strain $\delta u_{11}(\rho)$ (c) and potential variation $\delta\varphi(\rho)$ (d) across the cross-section of 180-degree cylindrical domain of radius $R = 10$ nm. The cylinder axes $z \equiv \tilde{x}_3$ coincides with [111] crystallographic direction. Parameters are the same as in **Figure 5**.

## 5. Summary and outcomes

Using Landau theory the analytical treatment of the anisotropic carrier accumulation by nominally uncharged domain walls in multiferroic $BiFeO_3$ was evolved. These theoretical results resolve some still outstanding issues in terms of conductivity in oxide ferroelectrics. Theoretical results are in a qualitative agreement with current-AFM experiments, namely uncharged 180-, 109- and 71-degree walls are indeed conducting in $BiFeO_3$.

The modeling results for anisotropic conductance around cylindrical or ring-type domain structures is also in a good agreement with recent experimental investigations [52], where anisotropic conductance was found around a ring domain structure and attributed to polarization discontinuities and associated migration of free carriers in the material to compensate the charge. Though the conductivity was found to be larger in the highly charged regions, significant conductivity could also be found in areas where the bound charge should have been close to zero.

Furthermore, it is now possible to explain some inconsistencies in the reported experimental data on this topic. For instance, the reported conductivity of 71° domain walls by Farokhipoor *et al* [13] in very thin BFO films can be reconciled by the fact that substrate-induced strain effects are likely larger in those films, and can therefore lead to significant conductivity through mechanisms outlined above. In contrast, thicker films with less dense domain structures show low conductivity at 71° walls [10, 52] presumably due to reduced contributions to conductivity that arise from the secondary strain-related effects.

The results presented here highlight that the effects of the angle-dependent electrostriction tensor and flexoelectric coupling-induced polarization cannot be neglected, and cause real, measurable increases to the static conductivity of domain walls in ferroelectrics. Lastly, these studies suggest that modulation of conduction by writing circular domains should be possible in a wide variety of standard ferroelectrics where charged domain walls are generally unstable due to high electrostatic energy cost, thus expanding the suite of materials for domain wall–based nanoelectronic applications.

**Acknowledgements**

A.N.M. and E.A.E. gratefully acknowledge multiple discussions, useful suggestions and critical remarks from Prof. A.K. Tagantsev and Dr. P.Yudin (EPFL, Lausanne) and Prof. V. Nagarajan (University of New South Wales, Kensington). Prof. Ying-Hao Chu (National Chiao Tung University, Hsinchu, Taiwan) is gratefully acknowledged for samples used to illustrate this study. E.A.E. and A.N.M. are thankful to NAS Ukraine and NSF-DMR-0908718 for support. R.K. V. acknowledges an overseas travel scholarship by the Australian Nanotechnology Network and the ARC Discovery Project Scheme. Research supported (S.V.K and P.M.) by the U.S. Department of Energy, Basic Energy Sciences, Materials Sciences and Engineering Division.

## Supplemental Materials. Tables

**Table S1a.** Dependence of the tensors and other coefficients on the wall rotation angle α of 180-degree domain wall in the **rhombohedral** phase

| Elastic compliances $\tilde{s}_{ij}$ in rotated frame $\{\tilde{x}_1, \tilde{x}_2, \tilde{x}_3\}$ | $\tilde{s}_{11} = \dfrac{2(s_{11}+s_{12})+s_{44}}{4}$, $\tilde{s}_{12} = \dfrac{2s_{11}+10s_{12}-s_{44}}{12}$, $\tilde{s}_{13} = \dfrac{2s_{11}+4s_{12}-s_{44}}{6}$, $\tilde{s}_{33} = \dfrac{s_{11}+2s_{12}+s_{44}}{3}$, $\tilde{s}_{66} = \dfrac{2(s_{11}-s_{12})+2s_{44}}{3}$, $\tilde{s}_{44} = \dfrac{4(s_{11}-s_{12})+s_{44}}{3}$, $\tilde{s}_{14} = \dfrac{\cos(3\alpha)}{3\sqrt{2}}(2(s_{11}-s_{12})-s_{44})$, $\tilde{s}_{15} = -\dfrac{\sin(3\alpha)}{3\sqrt{2}}(2(s_{11}-s_{12})-s_{44})$ |
|---|---|
| Electrostriction tensor components in $\{\tilde{x}_1, \tilde{x}_2, \tilde{x}_3\}$ | $\tilde{Q}_{11} = \dfrac{2(Q_{11}+Q_{12})+Q_{44}}{4}$, $\tilde{Q}_{12} = \dfrac{2Q_{11}+10Q_{12}-Q_{44}}{12}$, $\tilde{Q}_{13} = \dfrac{2Q_{11}+4Q_{12}-Q_{44}}{6}$, $\tilde{Q}_{33} = \dfrac{Q_{11}+2Q_{12}+Q_{44}}{3}$, $\tilde{Q}_{66} = \dfrac{2(Q_{11}-Q_{12})+2Q_{44}}{3}$, $\tilde{Q}_{44} = \dfrac{4(Q_{11}-Q_{12})+Q_{44}}{3}$, $\tilde{Q}_{14} = \dfrac{\cos(3\alpha)}{3\sqrt{2}}(2(Q_{11}-Q_{12})-Q_{44})$, $\tilde{Q}_{15} = -\dfrac{\sin(3\alpha)}{3\sqrt{2}}(2(Q_{11}-Q_{12})-Q_{44})$ |
| Combinations of the electrostriction coefficients $\vartheta_{ij}$ | $\vartheta_{11} = \tilde{Q}_{11} + \dfrac{\tilde{Q}_{12}(\tilde{s}_{14}^2\tilde{s}_{33}+(\tilde{s}_{12}\tilde{s}_{33}-\tilde{s}_{13}^2)\tilde{s}_{44})}{\tilde{s}_{14}^2\tilde{s}_{33}+(\tilde{s}_{13}^2-\tilde{s}_{11}\tilde{s}_{33})\tilde{s}_{44}} - \dfrac{\tilde{Q}_{13}\tilde{s}_{13}(2\tilde{s}_{14}^2+(\tilde{s}_{12}-\tilde{s}_{11})\tilde{s}_{44})}{\tilde{s}_{14}^2\tilde{s}_{33}+(\tilde{s}_{13}^2-\tilde{s}_{11}\tilde{s}_{33})\tilde{s}_{44}} - \dfrac{\tilde{Q}_{14}\tilde{s}_{14}(2\tilde{s}_{13}^2-(\tilde{s}_{11}+\tilde{s}_{12})\tilde{s}_{33})}{\tilde{s}_{14}^2\tilde{s}_{33}+(\tilde{s}_{13}^2-\tilde{s}_{11}\tilde{s}_{33})\tilde{s}_{44}}$  $\vartheta_{12} = \tilde{Q}_{12} + \dfrac{\tilde{Q}_{11}(\tilde{s}_{14}^2\tilde{s}_{33}+(\tilde{s}_{12}\tilde{s}_{33}-\tilde{s}_{13}^2)\tilde{s}_{44})}{\tilde{s}_{14}^2\tilde{s}_{33}+(\tilde{s}_{13}^2-\tilde{s}_{11}\tilde{s}_{33})\tilde{s}_{44}} - \dfrac{\tilde{Q}_{13}\tilde{s}_{13}(2\tilde{s}_{14}^2+(\tilde{s}_{12}-\tilde{s}_{11})\tilde{s}_{44})}{\tilde{s}_{14}^2\tilde{s}_{33}+(\tilde{s}_{13}^2-\tilde{s}_{11}\tilde{s}_{33})\tilde{s}_{44}} + \dfrac{\tilde{Q}_{14}\tilde{s}_{14}(2\tilde{s}_{13}^2-(\tilde{s}_{11}+\tilde{s}_{12})\tilde{s}_{33})}{\tilde{s}_{14}^2\tilde{s}_{33}+(\tilde{s}_{13}^2-\tilde{s}_{11}\tilde{s}_{33})\tilde{s}_{44}}$  $\vartheta_{13} = -\dfrac{(\tilde{Q}_{33}\tilde{s}_{13}-\tilde{Q}_{13}\tilde{s}_{33})(2\tilde{s}_{14}^2+(\tilde{s}_{12}-\tilde{s}_{11})\tilde{s}_{44})}{\tilde{s}_{14}^2\tilde{s}_{33}+(\tilde{s}_{13}^2-\tilde{s}_{11}\tilde{s}_{33})\tilde{s}_{44}}$, $\vartheta_{14} = \dfrac{(\tilde{Q}_{44}\tilde{s}_{14}-\tilde{Q}_{14}\tilde{s}_{44})((\tilde{s}_{11}+\tilde{s}_{12})\tilde{s}_{33}-2\tilde{s}_{13}^2)}{\tilde{s}_{14}^2\tilde{s}_{33}+(\tilde{s}_{13}^2-\tilde{s}_{11}\tilde{s}_{33})\tilde{s}_{44}}$,  $\vartheta_{15} = \dfrac{\vartheta_{16}}{2} = \dfrac{\tilde{Q}_{15}\tilde{s}_{44}(2\tilde{s}_{13}^2-(\tilde{s}_{11}+\tilde{s}_{12})\tilde{s}_{33})}{\tilde{s}_{14}^2\tilde{s}_{33}+(\tilde{s}_{13}^2-\tilde{s}_{11}\tilde{s}_{33})\tilde{s}_{44}}$ |
| | $\vartheta_{21} = \dfrac{((\tilde{Q}_{13}\tilde{s}_{13}-\tilde{Q}_{12}\tilde{s}_{33})\tilde{s}_{14}+\tilde{Q}_{14}(\tilde{s}_{13}^2-\tilde{s}_{11}\tilde{s}_{33}))\tilde{s}_{15}}{\tilde{s}_{14}^2\tilde{s}_{33}+(\tilde{s}_{13}^2-\tilde{s}_{11}\tilde{s}_{33})\tilde{s}_{44}}$,  $\vartheta_{22} = \dfrac{((\tilde{Q}_{13}\tilde{s}_{13}-\tilde{Q}_{11}\tilde{s}_{33})\tilde{s}_{14}-\tilde{Q}_{14}(\tilde{s}_{13}^2-\tilde{s}_{11}\tilde{s}_{33}))\tilde{s}_{15}}{\tilde{s}_{14}^2\tilde{s}_{33}+(\tilde{s}_{13}^2-\tilde{s}_{11}\tilde{s}_{33})\tilde{s}_{44}}$, $\vartheta_{23} = \dfrac{(\tilde{Q}_{33}\tilde{s}_{13}-\tilde{Q}_{13}\tilde{s}_{33})\tilde{s}_{14}\tilde{s}_{15}}{\tilde{s}_{14}^2\tilde{s}_{33}+(\tilde{s}_{13}^2-\tilde{s}_{11}\tilde{s}_{33})\tilde{s}_{44}}$,  $\vartheta_{24} = -\tilde{Q}_{15} + \dfrac{(\tilde{Q}_{14}\tilde{s}_{14}\tilde{s}_{33}+\tilde{Q}_{44}(\tilde{s}_{13}^2-\tilde{s}_{11}\tilde{s}_{33}))\tilde{s}_{15}}{\tilde{s}_{14}^2\tilde{s}_{33}+(\tilde{s}_{13}^2-\tilde{s}_{11}\tilde{s}_{33})\tilde{s}_{44}}$, $\vartheta_{25} = \tilde{Q}_{14} + \dfrac{\tilde{Q}_{15}\tilde{s}_{14}\tilde{s}_{15}\tilde{s}_{33}}{\tilde{s}_{14}^2\tilde{s}_{33}+(\tilde{s}_{13}^2-\tilde{s}_{11}\tilde{s}_{33})\tilde{s}_{44}}$,  $\vartheta_{26} = \dfrac{\tilde{Q}_{66}}{2} - \dfrac{2\tilde{Q}_{15}\tilde{s}_{15}(\tilde{s}_{13}^2-\tilde{s}_{11}\tilde{s}_{33})}{\tilde{s}_{14}^2\tilde{s}_{33}+(\tilde{s}_{13}^2-\tilde{s}_{11}\tilde{s}_{33})\tilde{s}_{44}}$ |
| | $\vartheta_{31} = \dfrac{\tilde{Q}_{15}}{2} - \dfrac{(\tilde{Q}_{14}\tilde{s}_{14}\tilde{s}_{33}+(\tilde{Q}_{12}\tilde{s}_{33}-\tilde{Q}_{13}\tilde{s}_{13})\tilde{s}_{44})\tilde{s}_{15}}{2(\tilde{s}_{14}^2\tilde{s}_{33}+(\tilde{s}_{13}^2-\tilde{s}_{11}\tilde{s}_{33})\tilde{s}_{44})}$,  $\vartheta_{32} = -\dfrac{\tilde{Q}_{15}}{2} + \dfrac{(\tilde{Q}_{14}\tilde{s}_{14}\tilde{s}_{33}+(\tilde{Q}_{13}\tilde{s}_{13}-\tilde{Q}_{11}\tilde{s}_{33})\tilde{s}_{44})\tilde{s}_{15}}{2(\tilde{s}_{14}^2\tilde{s}_{33}+(\tilde{s}_{13}^2-\tilde{s}_{11}\tilde{s}_{33})\tilde{s}_{44})}$, $\vartheta_{33} = \dfrac{(\tilde{Q}_{33}\tilde{s}_{13}-\tilde{Q}_{13}\tilde{s}_{33})\tilde{s}_{44}\tilde{s}_{15}}{2(\tilde{s}_{14}^2\tilde{s}_{33}+(\tilde{s}_{13}^2-\tilde{s}_{11}\tilde{s}_{33})\tilde{s}_{44})}$, |
| | $\vartheta_{34} = \dfrac{(\widetilde{Q}_{14}\widetilde{s}_{44} - \widetilde{Q}_{44}\widetilde{s}_{14})\widetilde{s}_{33}\widetilde{s}_{15}}{2(\widetilde{s}_{14}^2\widetilde{s}_{33} + (\widetilde{s}_{13}^2 - \widetilde{s}_{11}\widetilde{s}_{33})\widetilde{s}_{44})}$, $\vartheta_{35} = \dfrac{\widetilde{Q}_{44}}{2} + \dfrac{\widetilde{Q}_{15}\widetilde{s}_{15}\widetilde{s}_{33}\widetilde{s}_{44}}{2(\widetilde{s}_{14}^2\widetilde{s}_{33} + (\widetilde{s}_{13}^2 - \widetilde{s}_{11}\widetilde{s}_{33})\widetilde{s}_{44})}$, $\vartheta_{36} = \widetilde{Q}_{14} + \dfrac{\widetilde{Q}_{15}\widetilde{s}_{15}\widetilde{s}_{14}\widetilde{s}_{33}}{\widetilde{s}_{14}^2\widetilde{s}_{33} + (\widetilde{s}_{13}^2 - \widetilde{s}_{11}\widetilde{s}_{33})\widetilde{s}_{44}}$ |
| Flexoelectric tensor components in rotated frame $\{\widetilde{x}_1, \widetilde{x}_2, \widetilde{x}_3\}$ | $\widetilde{F}_{11} = \dfrac{F_{11} + F_{12} + F_{44}}{2}$, $\widetilde{F}_{12} = \dfrac{F_{11} + 5F_{12} - F_{44}}{6}$, $\widetilde{F}_{13} = \dfrac{F_{11} + 2F_{12} - F_{44}}{3}$, $\widetilde{F}_{33} = \dfrac{F_{11} + 2(F_{12} + F_{44})}{3}$, $\widetilde{F}_{66} = \dfrac{F_{11} - F_{12} + 2F_{44}}{3}$, $\widetilde{F}_{44} = \dfrac{2(F_{11} - F_{12}) + F_{44}}{3}$, $\widetilde{F}_{14} = \dfrac{\cos(3\alpha)}{3\sqrt{2}}(F_{11} - F_{12} - F_{44})$, $\widetilde{F}_{15} = -\dfrac{\sin(3\alpha)}{3\sqrt{2}}(F_{11} - F_{12} - F_{44})$ |
| Combinations of flexo-electric coefficients $\Psi_{ij}$ | $\Psi_{11} = -\widetilde{F}_{11} - \dfrac{\widetilde{F}_{12}(\widetilde{s}_{14}^2\widetilde{s}_{33} + (\widetilde{s}_{12}\widetilde{s}_{33} - \widetilde{s}_{13}^2)\widetilde{s}_{44})}{\widetilde{s}_{14}^2\widetilde{s}_{33} + (\widetilde{s}_{13}^2 - \widetilde{s}_{11}\widetilde{s}_{33})\widetilde{s}_{44}} + \dfrac{\widetilde{F}_{13}\widetilde{s}_{13}(2\widetilde{s}_{14}^2 + (\widetilde{s}_{12} - \widetilde{s}_{11})\widetilde{s}_{44})}{\widetilde{s}_{14}^2\widetilde{s}_{33} + (\widetilde{s}_{13}^2 - \widetilde{s}_{11}\widetilde{s}_{33})\widetilde{s}_{44}} + \dfrac{2\widetilde{F}_{14}\widetilde{s}_{14}(2\widetilde{s}_{13}^2 - (\widetilde{s}_{11} + \widetilde{s}_{12})\widetilde{s}_{33})}{\widetilde{s}_{14}^2\widetilde{s}_{33} + (\widetilde{s}_{13}^2 - \widetilde{s}_{11}\widetilde{s}_{33})\widetilde{s}_{44}}$ $\Psi_{12} = \dfrac{2\widetilde{F}_{15}\widetilde{s}_{14}((\widetilde{s}_{11} + \widetilde{s}_{12})\widetilde{s}_{33} - 2\widetilde{s}_{13}^2)}{\widetilde{s}_{14}^2\widetilde{s}_{33} + (\widetilde{s}_{13}^2 - \widetilde{s}_{11}\widetilde{s}_{33})\widetilde{s}_{44}}$, $\Psi_{13} = \dfrac{\widetilde{F}_{15}\widetilde{s}_{44}((\widetilde{s}_{11} + \widetilde{s}_{12})\widetilde{s}_{33} - 2\widetilde{s}_{13}^2)}{\widetilde{s}_{14}^2\widetilde{s}_{33} + (\widetilde{s}_{13}^2 - \widetilde{s}_{11}\widetilde{s}_{33})\widetilde{s}_{44}}$ $\Psi_{21} = \dfrac{\widetilde{s}_{15}\widetilde{s}_{14}(\widetilde{F}_{13}\widetilde{s}_{13} - \widetilde{F}_{12}\widetilde{s}_{33}) + 2\widetilde{F}_{14}\widetilde{s}_{15}(\widetilde{s}_{13}^2 - \widetilde{s}_{11}\widetilde{s}_{33})}{\widetilde{s}_{14}^2\widetilde{s}_{33} + (\widetilde{s}_{13}^2 - \widetilde{s}_{11}\widetilde{s}_{33})\widetilde{s}_{44}}$ $\Psi_{22} = -\dfrac{\widetilde{F}_{66}}{2} + \dfrac{2\widetilde{F}_{15}\widetilde{s}_{15}(\widetilde{s}_{13}^2 - \widetilde{s}_{11}\widetilde{s}_{33})}{\widetilde{s}_{14}^2\widetilde{s}_{33} + (\widetilde{s}_{13}^2 - \widetilde{s}_{11}\widetilde{s}_{33})\widetilde{s}_{44}}$, $\Psi_{23} = -\widetilde{F}_{14} - \dfrac{\widetilde{F}_{15}\widetilde{s}_{14}\widetilde{s}_{15}\widetilde{s}_{33}}{\widetilde{s}_{14}^2\widetilde{s}_{33} + (\widetilde{s}_{13}^2 - \widetilde{s}_{11}\widetilde{s}_{33})\widetilde{s}_{44}}$ $\Psi_{31} = -\widetilde{F}_{15} + \dfrac{2\widetilde{F}_{14}\widetilde{s}_{15}\widetilde{s}_{14}\widetilde{s}_{33} + \widetilde{s}_{15}\widetilde{s}_{44}(\widetilde{F}_{12}\widetilde{s}_{33} - \widetilde{F}_{13}\widetilde{s}_{13})}{2(\widetilde{s}_{14}^2\widetilde{s}_{33} + (\widetilde{s}_{13}^2 - \widetilde{s}_{11}\widetilde{s}_{33})\widetilde{s}_{44})}$, $\Psi_{32} = -\widetilde{F}_{14} - \dfrac{\widetilde{F}_{15}\widetilde{s}_{15}\widetilde{s}_{14}\widetilde{s}_{33}}{\widetilde{s}_{14}^2\widetilde{s}_{33} + (\widetilde{s}_{13}^2 - \widetilde{s}_{11}\widetilde{s}_{33})\widetilde{s}_{44}}$, $\Psi_{33} = -\dfrac{\widetilde{F}_{44}}{2} - \dfrac{\widetilde{F}_{15}\widetilde{s}_{15}\widetilde{s}_{33}\widetilde{s}_{44}}{2(\widetilde{s}_{14}^2\widetilde{s}_{33} + (\widetilde{s}_{13}^2 - \widetilde{s}_{11}\widetilde{s}_{33})\widetilde{s}_{44})}$ |

**Table S1b.** Dependence of the tensors and other coefficients on the wall rotation angle $\alpha$ for 109-degree walls and 71-degree walls in rhombohedral ferroelectric phase

| | |
|---|---|
| Elastic compliances $\widetilde{s}_{ij}$ in rotated frame $\{\widetilde{x}_1, \widetilde{x}_2, \widetilde{x}_3\}$ | $\widetilde{s}_{11} = s_{11} + \sin^2(2\alpha)\left(\dfrac{s_{44}}{4} - \dfrac{s_{11} - s_{12}}{2}\right)$, $\widetilde{s}_{12} = s_{12} - \sin^2(2\alpha)\left(\dfrac{s_{44}}{4} - \dfrac{s_{11} - s_{12}}{2}\right)$, $\widetilde{s}_{16} = -\widetilde{s}_{26} = -\dfrac{\sin(4\alpha)}{4}(s_{44} - 2(s_{11} - s_{12}))$, $\widetilde{s}_{66} = s_{44} - \sin^2(2\alpha)(s_{44} - 2(s_{11} - s_{12}))$ |
| Electrostriction tensor components in rotated frame $\{\widetilde{x}_1, \widetilde{x}_2, \widetilde{x}_3\}$ | $\widetilde{Q}_{11} = Q_{11} + \sin^2(2\alpha)\left(\dfrac{Q_{44}}{4} - \dfrac{Q_{11} - Q_{12}}{2}\right)$, $\widetilde{Q}_{12} = Q_{12} - \sin^2(2\alpha)\left(\dfrac{Q_{44}}{4} - \dfrac{Q_{11} - Q_{12}}{2}\right)$, $\widetilde{Q}_{16} = \widetilde{Q}_{61} = -\dfrac{\sin(4\alpha)}{4}(Q_{44} - 2(Q_{11} - Q_{12}))$, $\widetilde{Q}_{26} = \widetilde{Q}_{62} = \dfrac{\sin(4\alpha)}{4}(Q_{44} - 2(Q_{11} - Q_{12}))$, $\widetilde{Q}_{66} = Q_{44} - \sin^2(2\alpha)(Q_{44} - 2(Q_{11} - Q_{12}))$ |

| Combinations of the electrostriction coefficients $\vartheta_{ij}$ | $\vartheta_{11} = \widetilde{Q}_{11} - \dfrac{s_{12}\widetilde{s}_{11} - \widetilde{s}_{12}s_{12}}{s_{11}\widetilde{s}_{11} - s_{12}^2}Q_{12} - \dfrac{\widetilde{s}_{12}s_{11} - s_{12}^2}{s_{11}\widetilde{s}_{11} - s_{12}^2}\widetilde{Q}_{12}$, <br><br> $\vartheta_{12} = \widetilde{Q}_{12} - \dfrac{s_{12}\widetilde{s}_{11} - \widetilde{s}_{12}s_{12}}{s_{11}\widetilde{s}_{11} - s_{12}^2}Q_{12} - \dfrac{\widetilde{s}_{12}s_{11} - s_{12}^2}{s_{11}\widetilde{s}_{11} - s_{12}^2}\widetilde{Q}_{11}$, $\vartheta_{14} = \vartheta_{15} = 0$, <br><br> $\vartheta_{13} = \dfrac{s_{11}\widetilde{s}_{11} - \widetilde{s}_{12}s_{11}}{s_{11}\widetilde{s}_{11} - s_{12}^2}Q_{12} - \dfrac{s_{12}\widetilde{s}_{11} - \widetilde{s}_{12}s_{12}}{s_{11}\widetilde{s}_{11} - s_{12}^2}Q_{11}$, $\vartheta_{16} = \widetilde{Q}_{16} - \dfrac{\widetilde{s}_{12}s_{11} - s_{12}^2}{s_{11}\widetilde{s}_{11} - s_{12}^2}\widetilde{Q}_{26}$ |
|---|---|
| | $\vartheta_{21} = \widetilde{Q}_{61} - \dfrac{\widetilde{s}_{26}s_{11}\widetilde{Q}_{12}}{s_{11}\widetilde{s}_{11} - s_{12}^2} + \dfrac{\widetilde{s}_{26}s_{12}Q_{12}}{s_{11}\widetilde{s}_{11} - s_{12}^2}$, $\vartheta_{22} = \widetilde{Q}_{62} - \dfrac{\widetilde{s}_{26}s_{11}\widetilde{Q}_{11}}{s_{11}\widetilde{s}_{11} - s_{12}^2} + \dfrac{\widetilde{s}_{26}s_{12}Q_{12}}{s_{11}\widetilde{s}_{11} - s_{12}^2}$, <br><br> $\vartheta_{23} = \dfrac{\widetilde{s}_{26}(s_{12}Q_{11} - s_{11}Q_{12})}{s_{11}\widetilde{s}_{11} - s_{12}^2}$, $\vartheta_{24} = \vartheta_{25} = 0$, $\vartheta_{26} = \widetilde{Q}_{66} - \dfrac{\widetilde{s}_{26}s_{11}\widetilde{Q}_{26}}{s_{11}\widetilde{s}_{11} - s_{12}^2}$ |
| | $\vartheta_{31} = \vartheta_{32} = \vartheta_{33} = \vartheta_{34} = \vartheta_{36} = 0$, $\vartheta_{35} = \dfrac{Q_{44}}{2}$. |
| Flexoelectric tensor components in rotated frame $\{\widetilde{x}_1, \widetilde{x}_2, \widetilde{x}_3\}$ | $\widetilde{F}_{11} = F_{11} + \sin^2(2\alpha)\left(\dfrac{F_{44}}{2} - \dfrac{F_{11} - F_{12}}{2}\right)$, $\widetilde{F}_{12} = F_{12} - \sin^2(2\alpha)\left(\dfrac{F_{44}}{2} - \dfrac{F_{11} - F_{12}}{2}\right)$, <br><br> $\widetilde{F}_{16} = -\widetilde{F}_{26} = -\dfrac{\sin(4\alpha)}{4}(F_{44} - (F_{11} - F_{12}))$, $\widetilde{F}_{66} = F_{44} - \sin^2(2\alpha)(F_{44} - (F_{11} - F_{12}))$, <br><br> $\widetilde{F}_{61} = 2\widetilde{F}_{16}$, $\widetilde{F}_{62} = 2\widetilde{F}_{26}$ |
| Combinations of flexoelectric coefficients $\Psi_{ij}$ | $\Psi_{11} = -\widetilde{F}_{11} + \dfrac{(\widetilde{s}_{12}s_{11} - s_{12}^2)\widetilde{F}_{12} + (s_{12}\widetilde{s}_{11} - \widetilde{s}_{12}s_{12})F_{12}}{s_{11}\widetilde{s}_{11} - s_{12}^2}$, $\Psi_{12} = -\widetilde{F}_{16} + \dfrac{\widetilde{s}_{12}s_{11} - s_{12}^2}{s_{11}\widetilde{s}_{11} - s_{12}^2}\widetilde{F}_{26}$, <br><br> $\Psi_{13} = 0$ |
| | $\Psi_{21} = \dfrac{\widetilde{s}_{26}(s_{11}\widetilde{F}_{12} - s_{12}F_{12})}{s_{11}\widetilde{s}_{11} - s_{12}^2} - \widetilde{F}_{61}$, $\Psi_{22} = \dfrac{\widetilde{s}_{26}s_{11}\widetilde{F}_{26}}{s_{11}\widetilde{s}_{11} - s_{12}^2} - \widetilde{F}_{66}$, $\Psi_{23} = 0$ |
| | $\Psi_{31} = \Psi_{32} = 0$, $\Psi_{33} = -\dfrac{F_{44}}{2}$ |

Note, that expressions listed in the table are also valid for 180-degree domain wall in the **tetragonal** phase. Coordinate transformation is $\widetilde{x}_1 = x_1\cos\alpha - x_2\sin\alpha$, $\widetilde{x}_2 = x_1\sin\alpha + x_2\cos\alpha$ and $\widetilde{x}_3 = x_3$.

**Table S2.** Material parameters for bulk ferroelectric at room temperature 293°K

| coefficient | BiFeO$_3$ (collected and recalculated from Ref.[a]) | BaTiO$_3$ (collected and recalculated mainly from Ref. [b]) | PbZr$_{0.2}$Ti$_{0.8}$O$_3$ (collected and recalculated from Refs.[c, d]) |
|---|---|---|---|
| **Symmetry** | rhombohedral | tetragonal | tetragonal |
| $\varepsilon_b$ | 9 [e] | 7 [f] - 44 [g] | 5 [e] |
| $a_i$ (×10$^7$C$^{-2}$·mJ) | −7.53 | −2.94 | −14.84 |
| $a_{ij}$ (×10$^8$C$^-$ | $a_{11}$= 12, $a_{12}$= 2 | $a_{11}$= −6.71 | $a_{11}$= −0.305, $a_{12}$= 6.32 |
| $^4 \cdot m^5 J$) | | $a_{12}$= 3.23 | |
| $a_{ijk}$ ($\times 10^8 C^{-6} \cdot m^9 J$) | $a_{111}$=0, $a_{112}$=0, $a_{123}$=0 | $a_{111}$= 82.8, $a_{112}$=44.7, $a_{123}$=49.1 | $a_{111}$=2.475, $a_{112}$=9.68, $a_{123}$= −49.01 |
| $Q_{ij}$ ($C^{-2} \cdot m^4$) | $Q_{11}$=0.032, $Q_{12}$= −0.016, $Q_{44}$=0.010 | $Q_{11}$=0.11, $Q_{12}$= −0.043, $Q_{44}$=0.059 | $Q_{11}$=0.0814, $Q_{12}$= −0.0245, $Q_{44}$=0.0642 |
| $s_{ij}$ ($\times 10^{-12}$ Pa$^{-1}$) | $s_{11}$=5.29, $s_{12}$= −1.85, $s_{44}$=14.71 | $s_{11}$=8.3, $s_{12}$= −2.7, $s_{44}$=9.24 | $s_{11}$=8.2, $s_{12}$= −2.6, $s_{44}$=14.4 |
| $g_{ij}$ ($\times 10^{-10} C^{-2} m^3 J$) | $g_{11}$=4.0, $g_{44}$=2.0 * | $g_{11}$=5.1, $g_{12}$= −0.2, $g_{44}$= 0.2 [f] | $g_{11}$=2.0, $g_{44}$=1.0 * |
| $F_{ij}$ ($\times 10^{-11} C^{-1} m^3$) | $F_{11}$= −1.38, $F_{12}$= 0.67, $F_{44}$= 0.85 estimated using SrTiO$_3$ [h]<br>0.6 (estimated using Kogan model from $f_{ij} \sim$ 3.6V [i]) | ~100 (estimated from measurements of Ref. [j])<br>$F_{11}$= −2.46, $F_{12}$=0.48, $F_{44}$=0.05 (recalculated from Ref.[k] using $F_{\alpha\gamma}=f_{\alpha\beta}s_{\beta\gamma}$) | ~ 300 (estimated from measurements Ref. [l])<br><br>Separate values are not available |
| $\Xi_{ii}$ (eV) **Nonzero elements and values** | Nonzero elements $\Xi_{11}=\Xi_{22}=\Xi_{33}$<br>$\Xi_{12}=\Xi_{23}=\Xi_{13}$<br>Can be estimated from $\Xi_{11}^V + \Xi_{22}^V - \Xi_{11}^C - \Xi_{22}^C = 42$ [m]<br>$\Xi_{12}=\Xi_{23}=\Xi_{13}\approx 0$ | Nonzero elements $\Xi_{11}=\Xi_{22}$, $\Xi_{33}<0$<br><br>$\Xi_{11}+\Xi_{22}+\Xi_{33}=5$ Estimated from Ref.[n] separate values $\Xi_{ii}$ are unknown | Nonzero elements $\Xi_{11}=\Xi_{22}$, $\Xi_{33}<0$<br><br>for PbTiO$_3$ Ref.[o] $\Xi_{11}+\Xi_{22}+\Xi_{33}$= − 3.2 separate values $\Xi_{ii}$ are unknown |
| **Conductivity type** | $p$-type | $n$-type | $n$-type |

a) J. X. Zhang, Y. L. Li, Y. Wang, Z. K. Liu, L. Q. Chen, Y. H. Chu, F. Zavaliche, and R. Ramesh. J. Appl. Phys. 101, 114105 (2007).

b) A.J. Bell. J. Appl. Phys. **89**, 3907 (2001).

c) T.M. J. Haun, Z.Q. Zhuang, E. Furman, S.J. Jang and L.E. Cross. Ferroelectrics, Vol. 99, pp. 45-54 (1989).

d) N.A. Pertsev, V.G. Kukhar, H. Kohlstedt, and R. Waser, Phys. Rev. B **67**, 054107 (2003).

e) Estimation

f) J. Hlinka and P. Márton, Phys. Rev. B 74, 104104 (2006).

g) G. Rupprecht, R.O. Bell, Phys. Rev. 135, A748 (1964).

h) P. Zubko, G. Catalan, A. Buckley, P.R. L. Welche, J. F. Scott. Phys. Rev. Lett. **99**, 167601 (2007).

i) Sh. M. Kogan, *Sov. Phys.—Solid State* **5,** 2069 (1964)

j) W. Ma and L. E. Cross, Appl. Phys. Lett., **88**, 232902 (2006).

k) I. Ponomareva, A. K. Tagantsev, L. Bellaiche. Phys.Rev **B 85**, 104101 (2012).

l) W. Ma and L. E. Cross, Appl. Phys. Lett. 86, 072905 (2005).

m) Z. Fu, Z. G. Yin, N. F. Chen, X. W. Zhang, H. Zhang, Y. M. Bai, and J. L. Wu. Phys. Status Solidi RRL **6**, 1, 37–39 (2012)

## Supplemental Materials. Appendixes

### *S1. LGD-type equations of state for polarization components*

*S.1a. 109-degree walls and 71-degree walls in rhombohedral ferroelectric phase, as well as 180-degree domain wall in the tetragonal phase*

Equations of state for polarization components depending only on $\tilde{x}_1$ have the form [54]:

$$2a_1\tilde{P}_1 + 4\tilde{a}_{11}\tilde{P}_1^3 + 2\tilde{a}_{12}\tilde{P}_2^2\tilde{P}_1 + 2a_{12}\tilde{P}_3^2\tilde{P}_1 + \tilde{a}_{16}\tilde{P}_2(3\tilde{P}_1^2 - \tilde{P}_2^2) - \\ - \tilde{g}_{11}\frac{\partial^2 \tilde{P}_1}{\partial \tilde{x}_1^2} - \tilde{g}_{16}\frac{\partial^2 \tilde{P}_2}{\partial \tilde{x}_1^2} - 2(Q_{12}\tilde{\sigma}_3 + \tilde{Q}_{12}\tilde{\sigma}_2)\tilde{P}_1 - \tilde{Q}_{26}\tilde{\sigma}_2\tilde{P}_2 - \tilde{F}_{12}\frac{\partial \tilde{\sigma}_2}{\partial \tilde{x}_1} - F_{12}\frac{\partial \tilde{\sigma}_3}{\partial \tilde{x}_1} = \tilde{E}_1^d,$$
(S.1a)

$$2a_1\tilde{P}_2 + 4\tilde{a}_{11}\tilde{P}_2^3 + 2\tilde{a}_{12}\tilde{P}_1^2\tilde{P}_2 + 2a_{12}\tilde{P}_3^2\tilde{P}_2 + \tilde{a}_{16}\tilde{P}_1(\tilde{P}_1^2 - 3\tilde{P}_2^2) - \\ - \tilde{g}_{66}\frac{\partial^2 \tilde{P}_2}{\partial \tilde{x}_1^2} - \tilde{g}_{16}\frac{\partial^2 \tilde{P}_1}{\partial \tilde{x}_1^2} - 2(Q_{12}\sigma_3 + \tilde{Q}_{11}\sigma_2)\tilde{P}_2 - \tilde{Q}_{26}\tilde{\sigma}_2\tilde{P}_1 - \tilde{F}_{26}\frac{\partial \tilde{\sigma}_2}{\partial \tilde{x}_1} = 0$$
(S.1b)

$$2a_1\tilde{P}_3 + 4a_{11}\tilde{P}_3^3 + 2a_{12}(\tilde{P}_1^2 + \tilde{P}_2^2)\tilde{P}_3 - g_{44}\frac{\partial^2 \tilde{P}_3}{\partial \tilde{x}_1^2} - 2(Q_{11}\tilde{\sigma}_3 + Q_{12}\tilde{\sigma}_2)\tilde{P}_3 - Q_{44}\tilde{\sigma}_4\tilde{P}_2 = 0$$
(S.1c)

Depolarization field $\tilde{E}_1^d(\tilde{x}_1) \approx -\left(\tilde{P}_1(\tilde{x}_1) - \overline{\tilde{P}_1(\tilde{x}_1)}\right)/\varepsilon_0\varepsilon_b$ in the dielectric limit. Hereinafter tilda "~" defines vector or tensor components in the rotated frame, dash defines the spatial averaging. Coefficients in the rotated frame are:

$$\tilde{a}_{11} = a_{11} - \frac{2a_{11} - a_{12}}{4}\sin^2(2\alpha), \quad \tilde{a}_{12} = a_{12} + 3\frac{2a_{11} - a_{12}}{2}\sin^2(2\alpha), \quad \tilde{a}_{16} = \frac{2a_{11} - a_{12}}{2}\sin(4\alpha) \quad \text{(S.2a)}$$

$$\tilde{g}_{11} = g_{11} - \sin^2(2\alpha)\left(\frac{g_{11} - g_{12}}{2} - g_{44}\right), \quad \tilde{g}_{16} = \frac{\sin(4\alpha)}{2}\left(\frac{g_{11} - g_{12}}{2} - g_{44}\right), \quad \text{(S.2b)}$$

$$\tilde{g}_{66} = g_{44} + \sin^2(2\alpha)\left(\frac{g_{11} - g_{12}}{2} - g_{44}\right). \quad \text{(S.2c)}$$

Subscripts 1, 2 and 3 denote Cartesian coordinates *x, y, z* and Voigt's (matrix) notations are used: $a_{11} \equiv a_1$, $a_{1111} \equiv a_{11}$, $6a_{1122} \equiv a_{12}$, $g_{1111} \equiv g_{11}$, $g_{1122} \equiv g_{12}$, $g_{1212} \equiv g_{66}$, $Q_{1111} \equiv Q_{11}$, $Q_{1122} \equiv Q_{12}$, $4Q_{1212} \equiv Q_{44}$, $s_{1111} \equiv s_{11}$, $s_{1122} \equiv s_{12}$, $4s_{1212} \equiv s_{44}$, $F_{1111} \equiv F_{11}$, $F_{1122} \equiv F_{12}$, $2F_{1212} \equiv F_{44}$. Note that different factors (either "4", "2" or "1") in the definition of matrix notations with indices "44" are determined by the internal symmetry of tensors as well as by the symmetry of the corresponding physical properties tensors (see e.g. [55]). For the stress and strain tensors Voigt matrix notations have the view:

$$\sigma_{11} \equiv \sigma_1, \quad \sigma_{22} \equiv \sigma_2, \quad \sigma_{33} \equiv \sigma_3, \quad \sigma_{23} \equiv \sigma_4, \quad \sigma_{13} \equiv \sigma_5, \quad \sigma_{12} \equiv \sigma_6, \tag{S.3a}$$

$$u_{11} \equiv u_1, \quad u_{22} \equiv u_2, \quad u_{33} \equiv u_3, \quad u_{23} \equiv \frac{1}{2}u_4, \quad u_{13} \equiv \frac{1}{2}u_5, \quad u_{12} \equiv \frac{1}{2}u_6. \tag{S.3b}$$

### S.1b. 180-degree walls in rhombohedral ferroelectric phase

Equations of state for polarization components depending only on $\tilde{x}_1$ have the form

$$2a_1\tilde{P}_1 + 4\tilde{a}_{11}\tilde{P}_1^3 + 2\tilde{a}_{12}\tilde{P}_2^2\tilde{P}_1 + 2\tilde{a}_{13}\tilde{P}_3^2\tilde{P}_1 + 3\tilde{a}_{15}(\tilde{P}_1^2 - \tilde{P}_2^2)\tilde{P}_3 - 6\tilde{a}_{24}\tilde{P}_1\tilde{P}_2\tilde{P}_3 - \tilde{g}_{11}\frac{\partial^2 \tilde{P}_1}{\partial \tilde{x}_1^2} - \tilde{g}_{15}\frac{\partial^2 \tilde{P}_3}{\partial \tilde{x}_1^2} -$$
$$-2(\tilde{Q}_{12}\tilde{\sigma}_2 + \tilde{Q}_{13}\tilde{\sigma}_3 + \tilde{Q}_{14}\tilde{\sigma}_4)\tilde{P}_1 + 2\tilde{Q}_{15}\tilde{\sigma}_4\tilde{P}_2 + \tilde{Q}_{15}\tilde{\sigma}_2\tilde{P}_3 - \tilde{F}_{12}\frac{\partial \tilde{\sigma}_2}{\partial \tilde{x}_1} - \tilde{F}_{13}\frac{\partial \tilde{\sigma}_3}{\partial \tilde{x}_1} - 2\tilde{F}_{14}\frac{\partial \tilde{\sigma}_4}{\partial \tilde{x}_1} = \tilde{E}_1^d \tag{S.4a}$$

$$2a_1\tilde{P}_2 + 4\tilde{a}_{11}\tilde{P}_2^3 + 2\tilde{a}_{12}\tilde{P}_2^2\tilde{P}_1 + 2\tilde{a}_{13}\tilde{P}_3^2\tilde{P}_1 + 3\tilde{a}_{24}(-\tilde{P}_1^2 + \tilde{P}_2^2)\tilde{P}_3 - 6\tilde{a}_{15}\tilde{P}_1\tilde{P}_2\tilde{P}_3 - \tilde{g}_{66}\frac{\partial^2 \tilde{P}_2}{\partial \tilde{x}_1^2}$$
$$-\tilde{g}_{14}\frac{\partial^2 \tilde{P}_3}{\partial \tilde{x}_1^2} - 2(\tilde{Q}_{11}\tilde{\sigma}_2 + \tilde{Q}_{13}\tilde{\sigma}_3 - \tilde{Q}_{14}\tilde{\sigma}_4)\tilde{P}_2 + 2\tilde{Q}_{15}\tilde{\sigma}_4\tilde{P}_1 + (\tilde{Q}_{14}\tilde{\sigma}_2 - \tilde{Q}_{44}\tilde{\sigma}_4)\tilde{P}_3 + 2\tilde{F}_{15}\frac{\partial \tilde{\sigma}_4}{\partial \tilde{x}_1} = 0 \tag{S.4b}$$

$$2a_1\tilde{P}_3 + 4\tilde{a}_{33}\tilde{P}_3^3 + 2\tilde{a}_{13}(\tilde{P}_1^2 + \tilde{P}_2^2)\tilde{P}_3 + \tilde{a}_{15}(\tilde{P}_1^2 - 3\tilde{P}_2^2)\tilde{P}_1 + \tilde{a}_{24}(-3\tilde{P}_1^2 + \tilde{P}_2^2)\tilde{P}_2 - \tilde{g}_{44}\frac{\partial^2 \tilde{P}_3}{\partial \tilde{x}_1^2}$$
$$-\tilde{g}_{14}\frac{\partial^2 \tilde{P}_2}{\partial \tilde{x}_1^2} - \tilde{g}_{15}\frac{\partial^2 \tilde{P}_1}{\partial \tilde{x}_1^2} - 2(\tilde{Q}_{13}\tilde{\sigma}_2 + \tilde{Q}_{33}\tilde{\sigma}_3)\tilde{P}_3 + (\tilde{Q}_{14}\tilde{\sigma}_2 - \tilde{Q}_{44}\tilde{\sigma}_4)\tilde{P}_2 + \tilde{Q}_{15}\tilde{\sigma}_2\tilde{P}_1 + \tilde{F}_{15}\frac{\partial \tilde{\sigma}_2}{\partial \tilde{x}_1} = 0 \tag{S.4c}$$

Coefficients $a_1$, $a_{12}$ and $\tilde{a}_{11} = (2a_{11} + a_{12})/4$, $\tilde{a}_{12} = (2a_{11} + a_{12})/2$, $\tilde{a}_{13} = 2a_{11}$, $\tilde{a}_{33} = (a_{11} + a_{12})/3$, $\tilde{a}_{15} = -\sqrt{2}\sin(3\alpha)(2a_{11} - a_{12})/3$, $\tilde{a}_{24} = -\sqrt{2}\cos(3\alpha)(2a_{11} - a_{12})/3$ are the LGD-potential expansion coefficients. The gradient coefficients are $g_{ij}$ and $\tilde{g}_{11} = \frac{g_{11} + g_{12} + 2g_{44}}{2}$, $\tilde{g}_{44} = \frac{g_{11} - g_{12} + g_{44}}{3}$,

$\tilde{g}_{66} = \frac{g_{11} - g_{12} + 4g_{44}}{6}$, $\tilde{g}_{14} = \cos(3\alpha)\frac{g_{11} - g_{12} - 2g_{44}}{3\sqrt{2}}$, $\tilde{g}_{15} = -\sin(3\alpha)\frac{g_{11} - g_{12} - 2g_{44}}{3\sqrt{2}}$.

### S.2. Comments to solution of elastic fields

In the case of $\tilde{x}_1$-dependent solution, compatibility relation $e_{ikl}e_{jmn}(\partial^2 \tilde{u}_{ln}/\partial \tilde{x}_k \partial \tilde{x}_m) = 0$ leads to the conditions of constant strains $\tilde{u}_2 = const$, $\tilde{u}_3 = const$, $\tilde{u}_4 = const$, while general form dependences like $\tilde{u}_1 = \tilde{u}_1(\tilde{x}_1)$, $\tilde{u}_5 = \tilde{u}_5(\tilde{x}_1)$ and $\tilde{u}_6 = \tilde{u}_6(\tilde{x}_1)$ do not contradict to these relations. Mechanical equilibrium conditions $\partial \tilde{\sigma}_{ij}/\partial \tilde{x}_i = 0$ could be written as $\partial \tilde{\sigma}_1/\partial \tilde{x}_1 = 0$, $\partial \tilde{\sigma}_5/\partial \tilde{x}_1 = 0$, $\partial \tilde{\sigma}_6/\partial \tilde{x}_1 = 0$. Since $\tilde{\sigma}_{ij}(\tilde{x}_1 \to \pm\infty) = 0$, one obtains $\tilde{\sigma}_1 = \tilde{\sigma}_5 = \tilde{\sigma}_6 = 0$. For considered uncharged domain walls polarization-dependent elastic strains variations $\delta u_{ij}(r) = u_{ij}(z) - u_{ij}^S$ have the form:

$$\delta \tilde{u}_{22} = \delta \tilde{u}_{33} = \delta \tilde{u}_{23} = 0, \tag{S.5a}$$

$$\delta \tilde{u}_{i1} = \vartheta_{i1}\left(\tilde{P}_1^2 - \left(\tilde{P}_1^S\right)^2\right) + \vartheta_{i2}\left(\tilde{P}_2^2 - \left(\tilde{P}_2^S\right)^2\right) + \vartheta_{i3}\left(\tilde{P}_3^2 - \left(\tilde{P}_3^S\right)^2\right) + \vartheta_{i4}\left(\tilde{P}_2\tilde{P}_3 - \tilde{P}_2^S \tilde{P}_3^S\right)$$
$$+ \vartheta_{i5}\left(\tilde{P}_1\tilde{P}_3 - \tilde{P}_1^S \tilde{P}_3^S\right) + \vartheta_{i6}\left(\tilde{P}_1\tilde{P}_2 - \tilde{P}_1^S \tilde{P}_2^S\right) + \Psi_{i1}\frac{\partial \tilde{P}_1}{\partial \tilde{x}_1} + \Psi_{i2}\frac{\partial \tilde{P}_2}{\partial \tilde{x}_1} + \Psi_{i3}\frac{\partial \tilde{P}_3}{\partial \tilde{x}_1} \tag{S.5b}$$

Angle-dependent coefficients $\vartheta_{i1}$ and $\Psi_{i1}$ are listed in the **Tables S1a,b.**

### S.3. Approximate analytical solution for polarization components $\delta \tilde{P}_i(\tilde{x}_1)$ ($i = 1, 2$)

*S.3a. 109- and 71-degree domain walls in rhombohedral ferroelectric phase*

Neglecting the gradient terms and using perturbation approach on the small flexoelectric coupling coefficients, approximate solution of Eqs.(S.1) is possible. To derive the solution omit all terms proportional to the second powers of the flexoelectric coupling coefficients and their derivatives, i.e. in elastic stresses are omitted

$$\tilde{\sigma}_2 = \frac{s_{11}U_2 - s_{12}U_3}{s_{11}\tilde{s}_{11} - s_{12}^2}, \quad \tilde{\sigma}_3 = \frac{\tilde{s}_{11}U_3 - s_{12}U_2}{s_{11}\tilde{s}_{11} - s_{12}^2}, \quad \tilde{\sigma}_4 = \frac{Q_{44}\left(\tilde{P}_2^S \tilde{P}_3^S - \tilde{P}_2\tilde{P}_3\right)}{s_{44}}, \quad \tilde{\sigma}_1 = \tilde{\sigma}_5 = \tilde{\sigma}_6 = 0, \tag{S.6}$$

Functions $U_3$ and $U_2$ are introduced as

$$U_3 = Q_{11}\left(\left(\tilde{P}_3^S\right)^2 - \tilde{P}_3^2\right) + Q_{12}\left(\left(\tilde{P}_2^S\right)^2 + \left(\tilde{P}_1^S\right)^2 - \left(\tilde{P}_2^2 + \tilde{P}_1^2\right)\right) + F_{12}\frac{\partial \tilde{P}_1}{\partial \tilde{x}_1} \quad \text{and}$$

$$U_2 = \tilde{Q}_{11}\left(\left(\tilde{P}_2^S\right)^2 - \tilde{P}_2^2\right) + \tilde{Q}_{26}\left(\tilde{P}_1^S \tilde{P}_2^S - \tilde{P}_1\tilde{P}_2\right) + \tilde{Q}_{12}\left(\left(\tilde{P}_1^S\right)^2 - \tilde{P}_1^2\right) + Q_{12}\left(\left(\tilde{P}_3^S\right)^2 - \tilde{P}_3^2\right) + \tilde{F}_{12}\frac{\partial \tilde{P}_1}{\partial \tilde{x}_1} + \tilde{F}_{26}\frac{\partial \tilde{P}_2}{\partial \tilde{x}_1}.$$

After substitution of elastic stresses (S.6) in Eq.(S.1) the terms proportional to $F_{ij}^2$, $\tilde{F}_{kl}F_{ij}$ and $\tilde{F}_{ij}^2$ were neglected. As a result, polarization variation $\tilde{P}_3(\tilde{x}_1, F_{ij} \neq 0) - \tilde{P}_3(\tilde{x}_1, F_{ij} = 0) \equiv \delta \tilde{P}_3(\tilde{x}_1)$ appeared proportional to the second powers of the flexoelectric coupling coefficient. Expressions for the polarization variations $\delta \tilde{P}_i(\tilde{x}_1)$ induced by the flexoelectric and electrostriction couplings are listed below ($i = 1, 2$):

$$\delta \tilde{P}_2 \approx \frac{\tilde{Q}_{26}\tilde{\sigma}_2\tilde{P}_1^S + \tilde{F}_{26}(\partial \tilde{\sigma}_2/\partial \tilde{x}_1)}{2\left(a_1 + \tilde{a}_{12}\left(\tilde{P}_1^S\right)^2 + a_{12}\left(\tilde{P}_3^S\right)^2\right)} = 0 \tag{S.7a}$$

Since $\tilde{Q}_{26}(\alpha)=0$ and $F_{26}(\alpha)=0$ for the angles $\alpha=0$ and $\alpha=\pm\pi/4$. So the variation $\delta\tilde{P}_2(\tilde{x}_1)$ are absent for 109- and 71-degree uncharged domain walls.

$$\delta\tilde{P}_1 \approx \frac{\tilde{Q}_{26}\tilde{\sigma}_2\tilde{P}_2 + \tilde{F}_{12}(\partial\tilde{\sigma}_2/\partial\tilde{x}_1) + F_{12}(\partial\tilde{\sigma}_3/\partial\tilde{x}_1)}{2a_1 + 2\tilde{a}_{12}(\tilde{P}_2^S)^2 + 2a_{12}(\tilde{P}_3^S)^2 + (\varepsilon_0\varepsilon_b)^{-1}} = \frac{\tilde{F}_{12}(\partial\tilde{\sigma}_2/\partial\tilde{x}_1) + F_{12}(\partial\tilde{\sigma}_3/\partial\tilde{x}_1)}{2a_1 + 2\tilde{a}_{12}(\tilde{P}_2^S)^2 + 2a_{12}(\tilde{P}_3^S)^2 + (\varepsilon_0\varepsilon_b)^{-1}}$$

$$\approx \frac{\tilde{F}_{12}(\partial\tilde{P}_3^2/\partial\tilde{x}_1)\frac{Q_{11}s_{12}-s_{11}Q_{12}}{s_{11}\tilde{s}_{11}-s_{12}^2}}{2a_1 + 2\tilde{a}_{12}(\tilde{P}_2^S)^2 + 2a_{12}(\tilde{P}_3^S)^2 + (\varepsilon_0\varepsilon_b)^{-1}} + \frac{F_{12}(\partial\tilde{P}_3^2/\partial\tilde{x}_1)\frac{s_{12}Q_{12}-\tilde{s}_{11}Q_{11}}{s_{11}\tilde{s}_{11}-s_{12}^2}}{2a_1 + 2\tilde{a}_{12}(\tilde{P}_2^S)^2 + 2a_{12}(\tilde{P}_3^S)^2 + (\varepsilon_0\varepsilon_b)^{-1}} \quad \text{(S.7b)}$$

Using Eqs.(S.7), approximate analytical expressions for polarization components are

$$\tilde{P}_1 \approx \tilde{P}_1^S + \frac{f_1^Q}{\beta + (\varepsilon_0\varepsilon_b)^{-1}}\left(\frac{\partial\tilde{P}_3^2}{\partial\tilde{x}_1} + \frac{\partial\tilde{P}_2^2}{\partial\tilde{x}_1}\right), \quad \text{(S.8a)}$$

$$\tilde{P}_2 \approx \tilde{P}_2^S \tanh\left(\frac{\tilde{x}_1}{L_c}\right), \quad \text{(S.8b)}$$

$$\tilde{P}_3 \approx \tilde{P}_3^S \tanh\left(\frac{\tilde{x}_1}{L_c}\right) \quad \text{(S.8c)}$$

Where the constants $f_1^Q = \tilde{F}_{12}\frac{Q_{11}s_{12}-s_{11}Q_{12}}{s_{11}\tilde{s}_{11}-s_{12}^2} + F_{12}\frac{s_{12}Q_{12}-\tilde{s}_{11}Q_{11}}{s_{11}\tilde{s}_{11}-s_{12}^2}$ and $f_2^Q = \tilde{F}_{26}\frac{Q_{11}s_{12}-s_{11}Q_{12}}{s_{11}\tilde{s}_{11}-s_{12}^2}$ are introduced. Constant $\beta = 2(a_1 + \tilde{a}_{12}(\tilde{P}_2^S)^2 + a_{12}(\tilde{P}_3^S)^2)$, where coefficients $a_1$, $a_{12}$ and $\tilde{a}_{12}$ are the LGD-potential expansion coefficients; $\varepsilon_0\varepsilon_b$ is the background dielectric constant unrelated with soft mode.

### S.3b. 180-degree walls in rhombohedral ferroelectric phase

After substitution of corresponding elastic stresses

$$\tilde{\sigma}_2 = \frac{\tilde{s}_{44}(\tilde{s}_{33}U_2 - \tilde{s}_{13}U_3) + \tilde{s}_{14}\tilde{s}_{33}U_4}{(\tilde{s}_{11}\tilde{s}_{33}-\tilde{s}_{13}^2)\tilde{s}_{44} - \tilde{s}_{14}^2\tilde{s}_{33}}, \quad \tilde{\sigma}_3 = \frac{-\tilde{s}_{44}\tilde{s}_{13}U_2 + (\tilde{s}_{44}\tilde{s}_{11}-\tilde{s}_{14}^2)U_3 - \tilde{s}_{14}\tilde{s}_{13}U_4}{(\tilde{s}_{11}\tilde{s}_{33}-\tilde{s}_{13}^2)\tilde{s}_{44} - \tilde{s}_{14}^2\tilde{s}_{33}}, \quad \text{(S.9a)}$$

$$\tilde{\sigma}_4 = \frac{\tilde{s}_{14}(\tilde{s}_{33}U_2 - \tilde{s}_{13}U_3) + (\tilde{s}_{11}\tilde{s}_{33}-\tilde{s}_{13}^2)U_4}{(\tilde{s}_{11}\tilde{s}_{33}-\tilde{s}_{13}^2)\tilde{s}_{44} - \tilde{s}_{14}^2\tilde{s}_{33}}, \quad \tilde{\sigma}_1 = \tilde{\sigma}_5 = \tilde{\sigma}_6 = 0, \quad \text{(S.9b)}$$

Where the functions $U_3$ and $U_2$ are introduced as

$$U_2 = -\tilde{Q}_{12}\tilde{P}_1^2 - \tilde{Q}_{11}\tilde{P}_2^2 + \tilde{Q}_{13}\left((\tilde{P}_3^S)^2 - \tilde{P}_3^2\right) + \tilde{Q}_{14}\tilde{P}_2\tilde{P}_3 + \tilde{Q}_{15}\tilde{P}_1\tilde{P}_3 + \tilde{F}_{12}\frac{\partial\tilde{P}_1}{\partial\tilde{x}_1} - \tilde{F}_{15}\frac{\partial\tilde{P}_3}{\partial\tilde{x}_1}, \quad \text{(S.9c)}$$

$$U_3 = -\tilde{Q}_{13}(\tilde{P}_1^2 + \tilde{P}_2^2) + \tilde{Q}_{33}\left((\tilde{P}_3^S)^2 - \tilde{P}_3^2\right) + \tilde{F}_{13}\frac{\partial\tilde{P}_3}{\partial\tilde{x}_1}, \quad \text{(S.9d)}$$

and

$$U_4 = -\tilde{Q}_{44}\tilde{P}_2\tilde{P}_3 - \tilde{Q}_{14}(\tilde{P}_1^2 - \tilde{P}_2^2) + 2\tilde{Q}_{15}\tilde{P}_1\tilde{P}_2 + 2\tilde{F}_{14}\frac{\partial \tilde{P}_1}{\partial \tilde{x}_1} - 2\tilde{F}_{15}\frac{\partial \tilde{P}_2}{\partial \tilde{x}_1} \quad (S.9e)$$

After substitution of corresponding elastic stresses and cumbersome algebraic transformations, the evident form of Eqs.(S.4) is:

$$\left(2a_1 + \frac{1}{\varepsilon_0\varepsilon_b}\right)\tilde{P}_1 + \frac{(\tilde{Q}_{33}\tilde{s}_{13} - \tilde{Q}_{13}\tilde{s}_{33})\tilde{Q}_{15}\tilde{s}_{44}}{\tilde{s}_{14}^2\tilde{s}_{33} + (\tilde{s}_{13}^2 - \tilde{s}_{11}\tilde{s}_{33})\tilde{s}_{44}}\tilde{P}_3\left((\tilde{P}_3^S)^2 - \tilde{P}_3^2\right)$$

$$-\left(\tilde{g}_{15} + \frac{(\tilde{F}_{12}\tilde{s}_{44}\tilde{s}_{33} - \tilde{F}_{13}\tilde{s}_{13}\tilde{s}_{44} + 2\tilde{F}_{14}\tilde{s}_{14}\tilde{s}_{33})\tilde{F}_{15}}{\tilde{s}_{14}^2\tilde{s}_{33} + (\tilde{s}_{13}^2 - \tilde{s}_{11}\tilde{s}_{33})\tilde{s}_{44}}\right)\frac{\partial^2 \tilde{P}_3}{\partial \tilde{x}_1^2}$$

$$+\left(\frac{2\tilde{F}_{13}\tilde{Q}_{33}}{\tilde{s}_{33}} + \frac{2\left(\tilde{F}_{12}\tilde{s}_{44} + 2\tilde{F}_{14}\tilde{s}_{14} - \frac{\tilde{F}_{13}}{\tilde{s}_{33}}\tilde{s}_{13}\tilde{s}_{44}\right)(\tilde{Q}_{33}\tilde{s}_{13} - \tilde{Q}_{13}\tilde{s}_{33}) + \tilde{F}_{15}\tilde{Q}_{15}\tilde{s}_{44}\tilde{s}_{33}}{\tilde{s}_{14}^2\tilde{s}_{33} + (\tilde{s}_{13}^2 - \tilde{s}_{11}\tilde{s}_{33})\tilde{s}_{44}}\right)\tilde{P}_3\frac{\partial \tilde{P}_3}{\partial \tilde{x}_1}$$

$$-\left(\tilde{g}_{11} + \frac{4\tilde{F}_{14}^2(\tilde{s}_{13}^2 - \tilde{s}_{11}\tilde{s}_{33}) + 4\tilde{F}_{14}\tilde{s}_{14}(\tilde{F}_{13}\tilde{s}_{13} - \tilde{F}_{12}\tilde{s}_{33}) + \tilde{F}_{12}\tilde{s}_{44}(2\tilde{F}_{13}\tilde{s}_{13} - \tilde{F}_{12}\tilde{s}_{33}) + \tilde{F}_{13}^2(\tilde{s}_{14}^2 - \tilde{s}_{11}\tilde{s}_{44})}{\tilde{s}_{14}^2\tilde{s}_{33} + (\tilde{s}_{13}^2 - \tilde{s}_{11}\tilde{s}_{33})\tilde{s}_{44}}\right)\frac{\partial^2 \tilde{P}_1}{\partial \tilde{x}_1^2} \approx 0$$

(S.10a)

$$2\left\{a_1 + \frac{(\tilde{Q}_{33}\tilde{s}_{13} - \tilde{Q}_{13}\tilde{s}_{33})(\tilde{Q}_{11}\tilde{s}_{44} - \tilde{Q}_{14}\tilde{s}_{14}) - (\tilde{s}_{14}^2 - \tilde{s}_{11}\tilde{s}_{44})\tilde{Q}_{13}\tilde{Q}_{33} - \tilde{s}_{13}\tilde{s}_{44}\tilde{Q}_{13}^2}{\tilde{s}_{14}^2\tilde{s}_{33} + (\tilde{s}_{13}^2 - \tilde{s}_{11}\tilde{s}_{33})\tilde{s}_{44}}\left((\tilde{P}_3^S)^2 - \tilde{P}_3^2\right) + ..\right.$$
$$\left. + \left(\tilde{a}_{13} + \frac{2\tilde{s}_{14}\tilde{s}_{33}\tilde{Q}_{14}\tilde{Q}_{44} + (\tilde{s}_{13}^2 - \tilde{s}_{11}\tilde{s}_{33})\tilde{Q}_{44}^2 - \tilde{s}_{33}\tilde{s}_{44}\tilde{Q}_{14}^2}{2(\tilde{s}_{14}^2\tilde{s}_{33} + (\tilde{s}_{13}^2 - \tilde{s}_{11}\tilde{s}_{33})\tilde{s}_{44})}\right)\tilde{P}_3^2 + ..\right\}\tilde{P}_2 -$$

$$-\left(\tilde{g}_{66} + \frac{4\tilde{F}_{15}^2(\tilde{s}_{13}^2 - \tilde{s}_{11}\tilde{s}_{33})}{\tilde{s}_{14}^2\tilde{s}_{33} + (\tilde{s}_{13}^2 - \tilde{s}_{11}\tilde{s}_{33})\tilde{s}_{44}}\right)\frac{\partial^2 \tilde{P}_2}{\partial \tilde{x}_1^2} - \frac{(\tilde{Q}_{33}\tilde{s}_{13} - \tilde{Q}_{13}\tilde{s}_{33})(\tilde{Q}_{44}\tilde{s}_{14} - \tilde{Q}_{14}\tilde{s}_{44})}{\tilde{s}_{14}^2\tilde{s}_{33} + (\tilde{s}_{13}^2 - \tilde{s}_{11}\tilde{s}_{33})\tilde{s}_{44}}\tilde{P}_3\left((\tilde{P}_3^S)^2 - \tilde{P}_3^2\right)$$

$$-\left(\frac{4\tilde{s}_{14}(\tilde{Q}_{33}\tilde{s}_{13} - \tilde{Q}_{13}\tilde{s}_{33}) + \tilde{s}_{33}(\tilde{Q}_{44}\tilde{s}_{14} - \tilde{Q}_{14}\tilde{s}_{44})}{\tilde{s}_{14}^2\tilde{s}_{33} + (\tilde{s}_{13}^2 - \tilde{s}_{11}\tilde{s}_{33})\tilde{s}_{44}}\right)\tilde{F}_{15}\tilde{P}_3\frac{\partial \tilde{P}_3}{\partial \tilde{x}_1} - \left(\tilde{g}_{14} - \frac{2\tilde{F}_{15}^2\tilde{s}_{14}\tilde{s}_{33}}{\tilde{s}_{14}^2\tilde{s}_{33} + (\tilde{s}_{13}^2 - \tilde{s}_{11}\tilde{s}_{33})\tilde{s}_{44}}\right)\frac{\partial^2 \tilde{P}_3}{\partial \tilde{x}_1^2} \approx 0$$

(S.10b)

$$2a_1\tilde{P}_3 + 4\tilde{a}_{33}\tilde{P}_3^3 - \left(\tilde{g}_{44} - \frac{\tilde{F}_{15}^2\tilde{s}_{33}\tilde{s}_{44}}{\tilde{s}_{14}^2\tilde{s}_{33} + (\tilde{s}_{13}^2 - \tilde{s}_{11}\tilde{s}_{33})\tilde{s}_{44}}\right)\frac{\partial^2 \tilde{P}_3}{\partial \tilde{x}_1^2}$$

$$-2\frac{\tilde{Q}_{33}^2(\tilde{s}_{14}^2 - \tilde{s}_{11}\tilde{s}_{44}) + 2\tilde{Q}_{33}\tilde{Q}_{13}\tilde{s}_{13}\tilde{s}_{44} | -\tilde{Q}_{13}^2\tilde{s}_{33}\tilde{s}_{44}}{\tilde{s}_{14}^2\tilde{s}_{33} + (\tilde{s}_{13}^2 - \tilde{s}_{11}\tilde{s}_{33})\tilde{s}_{44}}\tilde{P}_3\left((\tilde{P}_3^S)^2 - \tilde{P}_3^2\right) + \quad (S.10c)$$

$$\frac{(\tilde{Q}_{33}\tilde{s}_{13} - \tilde{Q}_{13}\tilde{s}_{33})\tilde{Q}_{15}\tilde{s}_{44}}{\tilde{s}_{14}^2\tilde{s}_{33} + (\tilde{s}_{13}^2 - \tilde{s}_{11}\tilde{s}_{33})\tilde{s}_{44}}\tilde{P}_1\left((\tilde{P}_3^S)^2 - 3\tilde{P}_3^2\right) = 0$$

Note, that small $\tilde{P}_1$ component in Eqs.(S.10b,c) are neglected. Numerical simulations proved that the solution of Eq.(S.9c), its powers and derivatives can be well-approximated as:

$$\tilde{P}_3 \approx \tilde{P}_3^S \tanh\left(\frac{\tilde{x}_1}{L_c}\right) \quad (S.11a)$$

$$\frac{\partial \widetilde{P}_3}{\partial \widetilde{x}_1} \approx \frac{\widetilde{P}_3^S}{L_c} \cosh^{-2}\left(\frac{\widetilde{x}_1}{L_c}\right), \quad \left(\left(\widetilde{P}_3^S\right)^2 - \widetilde{P}_3^2\right) \approx \left(\widetilde{P}_3^S\right)^2 \cosh^{-2}\left(\frac{\widetilde{x}_1}{L_c}\right) \approx L_c \widetilde{P}_3^S \frac{\partial \widetilde{P}_3}{\partial \widetilde{x}_1} \tag{S.11b}$$

Neglecting the gradient terms is Eqs.(S.9a,b) and using Eq.(S.10b), approximate analytical expressions for polarization components $\widetilde{P}_1$ and $\widetilde{P}_2$ are

$$\widetilde{P}_1 \approx \frac{\varepsilon_0 \varepsilon_b f_1^Q}{1+2\beta\varepsilon_0\varepsilon_b} \frac{\partial \widetilde{P}_3^2}{\partial \widetilde{x}_1} + \frac{\varepsilon_0 \varepsilon_b q_1 \widetilde{P}_3}{1+2\beta\varepsilon_0\varepsilon_b}\left(\left(\widetilde{P}_3^S\right)^2 - \widetilde{P}_3^2\right) \approx \frac{\varepsilon_0 \varepsilon_b \psi_1^Q}{1+2\beta\varepsilon_0\varepsilon_b} \frac{\partial \widetilde{P}_3^2}{\partial \widetilde{x}_1}, \tag{S.12a}$$

$$\widetilde{P}_2 = \frac{f_2^Q}{2\beta} \frac{\partial \widetilde{P}_3^2}{\partial \widetilde{x}_1} + \frac{q_2}{2\beta} \widetilde{P}_3 \left(\left(\widetilde{P}_3^S\right)^2 - \widetilde{P}_3^2\right) \approx \frac{\psi_2^Q}{\beta} \frac{\partial \widetilde{P}_3^2}{\partial \widetilde{x}_1} \tag{S.12b}$$

Where the constants

$$\psi_1^Q = -\frac{1}{2}\left(\begin{array}{c} \dfrac{\left(\widetilde{Q}_{33}\widetilde{s}_{13} - \widetilde{Q}_{13}\widetilde{s}_{33}\right)\widetilde{Q}_{15}\widetilde{s}_{44}}{\widetilde{s}_{14}^2 \widetilde{s}_{33} + \left(\widetilde{s}_{13}^2 - \widetilde{s}_{11}\widetilde{s}_{33}\right)\widetilde{s}_{44}} L_c \widetilde{P}_3^S + \dfrac{2\widetilde{F}_{13}\widetilde{Q}_{33}}{\widetilde{s}_{33}} \\ + \dfrac{2\left(\widetilde{F}_{12}\widetilde{s}_{44} + 2\widetilde{F}_{14}\widetilde{s}_{14} - \dfrac{\widetilde{F}_{13}}{\widetilde{s}_{33}}\widetilde{s}_{13}\widetilde{s}_{44}\right)\left(\widetilde{Q}_{33}\widetilde{s}_{13} - \widetilde{Q}_{13}\widetilde{s}_{33}\right) + \widetilde{F}_{15}\widetilde{Q}_{15}\widetilde{s}_{44}\widetilde{s}_{33}}{\widetilde{s}_{14}^2 \widetilde{s}_{33} + \left(\widetilde{s}_{13}^2 - \widetilde{s}_{11}\widetilde{s}_{33}\right)\widetilde{s}_{44}} \end{array}\right) \tag{S.13a}$$

$$\psi_2^Q = \frac{\widetilde{F}_{15}\left(\begin{array}{c} 4\widetilde{s}_{14}\left(\widetilde{Q}_{33}\widetilde{s}_{13} - \widetilde{Q}_{13}\widetilde{s}_{33}\right) + \widetilde{s}_{33}\left(\widetilde{Q}_{44}\widetilde{s}_{14} - \widetilde{Q}_{14}\widetilde{s}_{44}\right) \\ + \left(\widetilde{Q}_{33}\widetilde{s}_{13} - \widetilde{Q}_{13}\widetilde{s}_{33}\right)\left(\widetilde{Q}_{44}\widetilde{s}_{14} - \widetilde{Q}_{14}\widetilde{s}_{44}\right)L_c \widetilde{P}_3^S \end{array}\right)}{2\left(\widetilde{s}_{14}^2 \widetilde{s}_{33} + \left(\widetilde{s}_{13}^2 - \widetilde{s}_{11}\widetilde{s}_{33}\right)\widetilde{s}_{44}\right)} \tag{S.13b}$$

$$\beta = a_1 + \frac{\left(\widetilde{Q}_{33}\widetilde{s}_{13} - \widetilde{Q}_{13}\widetilde{s}_{33}\right)\left(\widetilde{Q}_{11}\widetilde{s}_{44} - \widetilde{Q}_{14}\widetilde{s}_{14}\right) - \left(\widetilde{s}_{14}^2 - \widetilde{s}_{11}\widetilde{s}_{44}\right)\widetilde{Q}_{13}\widetilde{Q}_{33} - \widetilde{s}_{13}\widetilde{s}_{44}\widetilde{Q}_{13}^2}{\widetilde{s}_{14}^2 \widetilde{s}_{33} + \left(\widetilde{s}_{13}^2 - \widetilde{s}_{11}\widetilde{s}_{33}\right)\widetilde{s}_{44}}\left(\left(\widetilde{P}_3^S\right)^2 - \widetilde{P}_3^2\right) +$$
$$+ \left(\widetilde{a}_{13} + \frac{2\widetilde{s}_{14}\widetilde{s}_{33}\widetilde{Q}_{14}\widetilde{Q}_{44} + \left(\widetilde{s}_{13}^2 - \widetilde{s}_{11}\widetilde{s}_{33}\right)\widetilde{Q}_{44}^2 - \widetilde{s}_{33}\widetilde{s}_{44}\widetilde{Q}_{14}^2}{2\left(\widetilde{s}_{14}^2 \widetilde{s}_{33} + \left(\widetilde{s}_{13}^2 - \widetilde{s}_{11}\widetilde{s}_{33}\right)\widetilde{s}_{44}\right)}\right)\widetilde{P}_3^2 + ... \approx a_1 + \left(\widetilde{a}_{13} + \frac{\widetilde{Q}_{44}^2}{2\widetilde{s}_{44}}\right)\widetilde{P}_S^2 \tag{S.13c}$$